\documentclass[useAMS,usenatbib,referee]{biom}
\usepackage{amsmath}
\usepackage{graphicx}
\usepackage{amsfonts}
\usepackage{amssymb}
\usepackage{pgf}
\usepackage{xr}	% for cross-referencing sections from other articles
\newcommand{\Eps}{\boldsymbol{\mathcal E}}
\newcommand{\bs}{\boldsymbol}

\newcommand{\lt}{\left}
\newcommand{\rt}{\right}

\newcommand{\be}{\bs{e}}

\newcommand{\bR}{\bs{R}}
\newcommand{\bI}{\bs{I}}
\newcommand{\bU}{\bs{U}}

\newcommand{\bW}{\bs{W}}
\newcommand{\bY}{\bs{Y}}
\newcommand{\bX}{\bs{X}}

\newcommand{\bE}{\bs{E}}
\newcommand{\bH}{\bs{H}}
\newcommand{\bZ}{\bs{Z}}

\newcommand{\bbeta}{\bs{\beta}}
\newcommand{\omg}{\omega}
\newcommand{\del}{\delta}

\newcommand{\alp}{\alpha}

\newcommand{\Sig}{\Sigma}
\newcommand{\sig}{\sigma}

\newcommand{\eps}{\epsilon}
\newcommand{\beps}{\bs{\eps}}
\newcommand{\bSig}{\bs{\Sig}}
\newcommand{\bLam}{\bs{\Lambda}}

\renewcommand{\P}{\text{P}}
\renewcommand{\thesection}{\arabic{section}}

\newcommand{\Cov}{\text{Cov}}

\renewcommand{\theenumi}{(\roman{enumi})}%

\title[USAT: An Association Test for Multiple Phenotypes]{USAT: A Unified Score-based Association Test for
Multiple Phenotype-Genotype Analysis}

\author{Debashree Ray$^1$, James S Pankow$^2$ and Saonli Basu$^1$ \\
$^1$Division of Biostatistics, School of Public Health, University of Minnesota, U.S.A.\\
$^2$Division of Epidemiology \& Community Health, School of Public Health, University of Minnesota, U.S.A.
}

\begin{document}

\label{firstpage}
\clearpage

%  put the summary for your paper here

\begin{abstract}

Genome-wide Association Studies (GWASs) for complex diseases often collect data on multiple correlated endo-phenotypes. Multivariate analysis of these correlated phenotypes can improve the power to detect genetic variants. Multivariate analysis of variance (MANOVA) can perform such association analysis at a GWAS level, but the behavior of MANOVA under different trait models has not been carefully investigated. In this paper, we show that MANOVA is generally very powerful for detecting association but there are situations, such as when a genetic variant is associated with all the traits, where MANOVA may not have any detection power. We investigate the behavior of MANOVA, both theoretically and using simulations, and derive the conditions where MANOVA loses power. 
Based on our findings, we propose a unified score-based test statistic USAT that can perform better than MANOVA in such situations 
and nearly as well as MANOVA elsewhere. Our proposed test reports an approximate asymptotic p-value for association and is computationally very efficient to implement at a GWAS level. We have studied through extensive simulation the performance of USAT, MANOVA and other existing approaches and demonstrated  the advantage of using the USAT approach to detect association between a genetic variant and multivariate phenotypes.
We applied USAT to data from three correlated traits collected on $5,816$ Caucasian individuals from the Atherosclerosis Risk in Communities (ARIC, \cite{ARIC89}) Study and detected some interesting associations.\\ 
\end{abstract}

\begin{keywords}
GWAS; MANOVA; Multiple correlated phenotypes; Multivariate analysis; Score-based test
\end{keywords}

\maketitle

\section{Introduction}

In the study of a complex disease, data on several correlated endo-phenotypes are often collected to get a better understanding of the disease. 
For example, in the study of thrombosis, the intermediate correlated phenotypes such as Factor VII, VIII, IX, XI, XII, and von Willebrand factor influence greatly the risk of developing thrombosis \citep{Souto00,Germain11}. An epidemiologic study on type 2 diabetes (T2D) typically collects data on  
a number of risk factors and diabetes-related quantitative traits. The standard approach to analyze these phenotypes is to perform single-trait analyses separately and report the findings for individual trait.

\cite{Sluis13} demonstrated several alternative models which would benefit from a joint analysis.
\cite{Blair13} illustrated the comorbidity between Mendelian disorders and different complex disorders, which indicates that there may be common genetic variants affecting several of these complex traits. 
Recently, many articles advocating joint analysis over univariate analysis of multiple correlated traits
\citep[and references therein]{Ferreira09,Zhang09,Korte12,OReilly12,Stephens13,Aschard14,Galesloot14,Zhou14,Ried14}
have been published that illustrate the benefits of jointly analyzing these correlated traits to improve the power of detection of genetic variants.
Moreover this joint analysis could reveal some pleiotropic genes involved in the biological development of the disease. 

Few approaches have been developed to perform association analysis with multivariate traits at a GWAS level.
 \cite{OReilly12} proposed MultiPhen to detect association between multivariate traits and a  single nucleotide polymorphism (SNP) with unrelated individuals. MultiPhen uses ordinal regression to regress a SNP on a collection of phenotypes and tests whether all regression parameters corresponding to the phenotypes in the model are significantly different from zero. 
It can accommodate both binary and continuous traits but may suffer from lack of power when a SNP is associated with all the highly correlated traits.
\cite{Sluis13} proposed Trait-based Association Test (TATES) for testing association between multiple traits and multiple SNPs using extended Simes procedure on the p-values derived from univariate trait and single SNP association analysis. Even when the phenotypes are strongly correlated, TATES gives appropriate type I error for varying minor allele frequency (m.a.f.). It may have low power when a SNP affects only a few of the strongly correlated traits.  
\cite{Maity12} proposed a kernel machine method for unrelated individuals for joint analysis of multimarker effects on multiple traits. Kernel machine is a powerful dimension-reduction tool that can accommodate linear/non-linear effects of multiple SNPs. Their test for association between multiple SNPs and the phenotypes is equivalent to testing the variance components in a multivariate linear mixed model (mvLMM). Implementation of this approach requires parametric bootstrapping to estimate the distribution of the test statistic and could be computationally intensive at a GWAS level. 
\cite{Korte12,Zhou12} implemented mvLMM for GWAS.
\cite{Zhou14} explored efficient algorithms for mvLMM in a GWAS setting.

Recently, data reduction methods such as principal component analysis (PCA) and canonical correlation analysis (CCA) are being explored to perform multivariate association analysis \citep{Tang12, Basu13, Aschard14}. The advantage of using CCA to perform gene-based tests on multivariate phenotypes has been elaborately discussed in \cite{Tang12, Basu13}. Previously, \cite{Ferreira09} proposed a 
multivariate test of association based on CCA to simultaneously test the association between a single SNP and multiple phenotypes.  Their CCA approach is equivalent to multivariate analysis of variance (MANOVA) or more generally the Wilk's lambda test in multivariate multiple linear regression (MMLR) approach \citep{Muller84}. \cite{Basu13} extended the MANOVA to family data. 
Both  \cite{OReilly12} and \cite{Sluis13} found significantly high power for MANOVA when a subset of traits was associated with the causal variant or gene. One major advantage of MANOVA is that it can easily be extended to incorporate multiple phenotypes as well as multiple SNPs (such as a gene). Moreover other covariates can easily be incorporated in the model. 

In this paper, we explore the performance of MANOVA to detect multi-trait association under various alternative trait models.  Our simulation studies consider a single marker to investigate the properties of MANOVA. Further, we theoretically justify the behavior of MANOVA and provide a geometrical explanation as well.
We demonstrate that MANOVA may lose significant power when the genetic marker is associated with all the traits and any test that does not consider the within trait correlation can have more power in such a  situation. 
Utilizing these findings, we propose a novel unified score-based association test (USAT) that maintains good power under various alternative trait models and performs significantly better than MANOVA when all the traits are associated.

This paper evolves as follows. Section \ref{sec:method} describes some popular existing methods for doing association analysis using multiple phenotypes. More specifically, section \ref{sec:univ} describes the univariate methods that completely ignore trait correlations, section \ref{sec:ssu} describes a method that accounts for the within trait correlation only through the distribution of the test statistic while section \ref{sec:multi} describes a multivariate method that directly incorporates the trait correlation structure. 
Section \ref{sec:manprop} theoretically and geometrically justifies some aspects of the behavior of MANOVA, for $K$ traits and a single SNP, in situations that commonly arise in such genetic studies. 
Section \ref{sec:usat} introduces our unified approach USAT for association analysis using multiple traits and a single marker for unrelated individuals. 
Section \ref{sec:result} illustrates a comparison of different existing approaches and USAT using simulated data and a real dataset.
Section \ref{sec:discuss} concludes this article with a short summary and discussion.

\section{Methods} \label{sec:method}

Consider $K$ correlated traits $Y_1, Y_2, \ldots, Y_K$ in $n$ unrelated individuals. Let $\bY_k$ be the $n\times1$ vector of $k$-th trait and $\bY$ be the $n\times K$ matrix of traits for all individuals.  Consider a GWAS setting with data on a large number $p$ $(\gg n)$ of genetic variants.
We are interested in testing the association of a single SNP with the $K$ correlated traits.
For a given SNP, let $X_i$ be the number of copies of minor alleles $(0,1 \text{ or } 2)$ for $i$-th individual and $\bX$ be the $n\times1$ vector of genotypes for all samples. Without loss of generality, it is assumed that the phenotype matrix $\bY$ and the genotype vector $\bX$ are centered but not standardized. 

Due to the correlatedness of the traits, a standard approach would be to consider an MMLR model for the association test of $K$ traits and the SNP:
\begin{equation}
\bY_{n\times K} = \bX_{n\times1} \bbeta_{1\times K}' + \Eps_{n\times K} \label{mmlr}
\end{equation}
where $\bbeta'=(\beta_1,...,\beta_K)$ is the vector of fixed unknown genetic effects corresponding to the $K$ correlated traits, and 
$\Eps$ is the matrix of random errors. 
For testing that the SNP is not associated with any of the $K$ traits, the null hypothesis of interest is $H_0: \bbeta=\bs{0}$.

In the MMLR model \eqref{mmlr}, each row of $\Eps$ is i.i.d. with mean $\bs{0}_{K\times1}$ and variance $\bSig_{K\times K}$. 
In particular, $\Eps$ may be assumed to be an $n\times K$ normal data matrix from $N_K(\bs{0},\bSig)$, where $\bSig$ is a positive definite (p.d.) matrix representing residual covariance among the traits. The likelihood ratio test (LRT) of $H_0$ based on the MMLR model with matrix normal errors is equivalent to MANOVA \citep{Muller84,Yang12}.
One may consider a further partition of $\Eps$ to arrive at mvLMM:
$$\bY_{n\times K}= \bX_{n\times1} \bbeta'_{1\times K} + \bW_{n\times K} + \beps_{n\times K}$$ 
where $\bW$ is a matrix of random effects representing heritable component of the phenotypes, and
$\beps$ is the matrix of errors characterizing random variation arising from unmeasured sources. 
In recent times, mvLMMs have been recognized as powerful tools for testing $H_0$. mvLMM can not only control population structure and other confounding factors, but also accounts for relatedness among multiple traits. Association tests based on mvLMM can be computationally challenging and many efficient algorithms have been developed to this end \citep{Yang11,Korte12,Zhou14}.

Apart from multivariate models, one may use marginal models for such an association test. Although marginal modeling effectively assumes the traits to be uncorrelated, approaches based on marginal models are often computationally faster and easier to implement.
The marginal model for testing association of a SNP with $k$-th trait is given by
\begin{equation}
\bY_k = \beta_{M,k} \bX + \be_k, \: \be_k \sim N(\bs{0}, \sig^2 \bI_n), \: \:k=1,2,...,K	 \label{marginal}
\end{equation}
$\beta_{M,k}$ is the $k$-th genetic effect in the marginal model.
 For the $k$-th marginal model, our null hypothesis is $H_{0,k}: \beta_{M,k}=0$. In order to carry out the simultaneous test $H_0$, one still needs to devise an approach to combine the results from the marginal tests $H_{0,k},\:k=1,2,...,K$.

Broadly, the different statistical approaches for testing our global null hypothesis of no association can be classified into three categories: (1) tests that completely ignore the within trait correlation; (2) tests that incorporate within trait correlation only in deriving the distribution of the test statistic; and (3) tests that incorporate the within trait correlation directly in deriving the test statistic. We compare through extensive simulation studies these three broad approaches and discuss their advantages and shortcomings under various alternative trait models.

\subsection{Combination Tests that completely ignore within trait correlation} \label{sec:univ}
This category of tests considers separate regression models for the $K$ traits (i.e., $K$ univariate analyses), thereby treating the traits as uncorrelated. 
Let $p_k$ be the p-value for testing $H_{0,k}$ based on the $k$-th marginal model in \eqref{marginal}. This class of tests proposes several approaches of combining the p-values $p_1,...,p_K$ for testing our global null hypothesis $H_0 = \cap_{k=1}^K H_{0,k}$.

\subsubsection{Fisher's Test}
Fisher's method \citep{Fisher} involves combining the logarithmic transformation of the p-values $p_1,...,p_K$.
The test statistic is $-2 \text{log}_e \sum_{k=1}^K p_k$, which under $H_0$ 
and the assumption of independent tests, has a $\chi^2_{2K}$ distribution. 
In the presence of strong correlation among traits, inflated type-I error is observed (`anti-conservative').

\subsubsection{minP Test}
The minP test statistic is based on the minimum of adjusted p-values, where adjustment is usually done by Bonferroni's method to take care of multiple-testing issue. It is given by $p_{\min} = \text{min}_{k=1}^K {K p_k}$.  
Under $H_0$ and the assumption of independence among the phenotypes, 
$p_{\min}$ is distributed as the minimum of independent $U(0,1)$ variables.
In the presence of correlation structure, this test can be conservative.
To take care of this conservativeness, \cite{Sluis13} proposed TATES which combines p-values from univariate analyses while correcting for the relatedness among the phenotypes.

\subsection{Test that incorporates trait correlation only through distribution} \label{sec:ssu}
This category of tests does not explicitly consider the trait correlation in the test statistics. The correlation is taken into account in finding the true null distribution of the test statistic due to which the statistic maintains proper type I error. 
A notable test in this category is the Sum of Squared Score (SSU) test as outlined by \cite{Yang12}, an extension of the SSU test for association of multiple SNPs with a single trait proposed by \cite{Pan09}.

\subsubsection{SSU Test}
SSU is a score-based test where the score vector is derived from the marginal normal models in equation \eqref{marginal}.
Under the global null $H_0$, the $K\times1$ vector of marginal scores is given by
$$\bU_M =\frac{1}{\hat\sig_0^2} \bY'\bX
$$
where $\hat\sig_0^2 = \frac{1}{K(n-1)}\sum_{i=1}^n\sum_{k=1}^K Y_{ik}^2$ is the MLE of $\sig^2$ in equation \eqref{marginal} under the null.
 The SSU test statistic is $T_{S}=\bU_M'\bU_M$,
which has an approximate asymptotic scaled and shifted chi-squared distribution
$a\chi^2_d + b$ \citep{Zhang05} under $H_0$. 
 The distributional parameters are determined as 
\begin{equation} \label{ssupar}
a=\frac{\sum c_k^3}{\sum c_k^2}, \:\: 
b=\sum c_k  - \frac{(\sum c_k^2)^2}{\sum c_k^3}, \:\:
d = \frac{(\sum c_k^2)^3}{(\sum c_k^3)^2}
\end{equation}
 where $\{c_k\}_{k=1}^K$ are the ordered eigenvalues of 
$\Cov(\bU_M) = \bX'\bX \bY'\bY/(n\hat\sig_0^2)$.

An important aspect of the SSU test is that the test statistic does not incorporate the trait covariance structure. 
Notice that, according to equation \eqref{ssupar}, $\Cov(\bU_M)$ contains information on within trait correlations and  is used in deriving the distribution of the statistic.
If $\bU$ be the score vector from MMLR model \eqref{mmlr} under $H_0$, a test statistic of the form $\bU'\bU$ will not be an SSU type test since the within trait covariance matrix is incorporated in $\bU$.

\subsection{Multivariate Test that incorporates within trait correlation directly in the test statistic} \label{sec:multi}

This class of tests explicitly incorporates the within trait correlation structure in the test statistics as well as in finding their distributions.

\subsubsection{MANOVA} \label{sec:cca}
Consider the MMLR model in equation \eqref{mmlr}. Assume each row of $\Eps$ to be i.i.d. $N_K(\bs{0},\bSig)$.
The log-likelihood for the data matrix $\bY$ is given by
\begin{equation}
l(\bbeta, \bSig)  = -\frac{1}{2}n \log|2\pi\bSig| - \frac{1}{2} \text{tr} \lt\{ \bSig^{-1}  (\bY-\bX\bbeta')' (\bY-\bX\bbeta') \rt\} \label{loglik}
\end{equation}
For testing $H_0$, the LRT is equivalent to the MANOVA test statistic (Wilk's Lambda), which is the ratio of generalized variances 
${|\bE|}/{|\bH+\bE|}$. Here $\bH$ is the hypothesis sum of squares and cross product (SSCP) matrix and $\bE$ is the error SSCP matrix. 
The explicit forms of these SSCP matrices in terms of phenotype and genotype data are $\bH=\hat\bbeta(\bX'\bX)\hat\bbeta'$ and $\bE=\bY'\bY-\hat\bbeta(\bX'\bX)\hat\bbeta'$, where $\hat\bbeta=\bY'\bX(\bX'\bX)^{-1}$ is the MLE of $\bbeta$.
Thus, $\bH$ is calculated as the covariance matrix of the fitted values, and $\bE$ is calculated as the covariance matrix of the residuals of the model. Under $H_0$,
$-2 \log\bLam = -n \log \lt( {|\bE|}/{|\bH+\bE|} \rt)$ has an approximate asymptotic $\chi^2_K$ distribution. 

Another such multivariate approach is MultiPhen where the genotype is modeled as ordinal using a proportional odds regression model. \cite{OReilly12} empirically showed that for a single SNP, MultiPhen's performance is similar to MANOVA.

\subsection{MANOVA and its behavior} \label{sec:manprop}

A major challenge in multivariate disease-related trait analysis is the lack of a test that is uniformly most powerful under different patterns/levels of association and different within trait correlation structures. 
The association tests which do not consider within trait correlation at all are either `conservative' or `anti-conservative'. 
Our simulation studies with exchangeable correlation structure show that MANOVA generally has better performance but loses significant power when within trait correlation is high and is in the same direction as all the genetic effects. For a moderate number of traits,  MANOVA may fail to detect pleiotropy (phenomenon where a single genetic variant affects all the traits) even at low within trait correlations (refer sections \ref{sec:sim1}, \ref{sec:sim2}).

The following theorems provide conditions under which MANOVA loses power when a SNP is associated with all $K$ correlated traits. We assume a compound symmetry (CS) residual correlation structure. Theorem proofs are provided in \ref{S-app1}.

\begin{theorem} \label{Thm3}
Consider the MMLR model $\bY_{n\times K} = \bX_{n\times1} \bbeta_{1\times K}' + \Eps_{n\times K}$ with 
$vec(\Eps)\sim N_{nK}(\bs{0}, \bI_n\otimes\bSig)$, $\bSig = \sig^2 \lt( (1-\rho)\bI_K + \rho \bs{1}\bs{1}' \rt)$, 
$\sig^2>0$, $\rho$ $(>0)$ is the within trait correlation such that $\bSig$ is a p.d. matrix,
and $\bbeta'=(\beta_1,...,\beta_K)$ is the vector of genetic effects.
%and $\bbeta_0$ is the vector of intercepts. 
%Assume that $\bX=(X_1,...,X_n)'$, where each $X_i$ can take values $0,1,2$ with probabilities $(1-f)^2, 2f(1-f), f^2$ respectively.
%$\E(X_i)=f$, $\Var(X_i)=2f(1-f) \: \forall \: i=1,2,...,n$. 
Assume that the genetic effects of the associated traits are equal in size and in positive direction. Consider two scenarios of association: `partial association' (when the SNP is associated with $u$ $(<K)$ traits), and `complete association' (when all $K$ traits are associated).
For testing $H_0: \bbeta=\bs{0}$, the power of MANOVA under partial association will be asymptotically more than that under complete association 
if $\frac{u}{K} > \frac{1-\rho}{1+(K-u-1)\rho} 
= \frac{\text{2nd eigenvalue of }\bSig_{K-u}}{\text{1st eigenvalue of }\bSig_{K-u}}$. Here $\bSig_{K-u}$ is the CS residual covariance matrix of the $K-u$ truly unassociated traits. 
\end{theorem}
 
%Note that for $u=1$ and $K=2$ correlated traits, theorem \ref{Thm3} boils down to corollary \ref{cor1}.

For $K=2$ traits, Theorem \ref{Thm3} can be generalized further to encompass genetic effects in opposite direction, and negative within trait correlation. 

\begin{theorem} \label{Thm2}
%\begin{lemma} \label{Thm2}
Consider the MMLR model in Theorem \ref{Thm3} with $K=2$ traits. The genetic effects of the associated traits may or may not be equal in size or in same direction. The within trait correlation $\rho$ may or may not be positive. For testing $H_0: \beta_1=\beta_2=0$, the power of MANOVA when only one trait is associated is asymptotically more than when both traits are associated if $0<\beta_2<2\rho\beta_1$ or $0>\beta_2>2\rho\beta_1$. 
\end{theorem}
%\end{lemma}

\begin{corollary} \label{cor}
In particular, let us assume that the genetics effects of the associated traits are equal in size. That is, $|\beta_1|=|\beta_2|$ when the SNP is associated with both the correlated traits. Asymptotically, the power of MANOVA under $H_{a1}:\beta_1>0,\beta_2=0$ will exceed the power of MANOVA under
\begin{enumerate}[ i{)}]
\item $H_{a2,1}:\beta_1=\beta_2>0$ when $\rho>1/2$;
\item $H_{a2,2}:\beta_1=-\beta_2>0$ when $\rho<-1/2$.
\end{enumerate}
\end{corollary}

%\subsubsection{A geometric explanation} \label{sec:geom}

\begin{figure}[t!]
	\begin{center}
%	\begin{subfigure}[b]{0.3\textwidth}
	\includegraphics[width=3in]{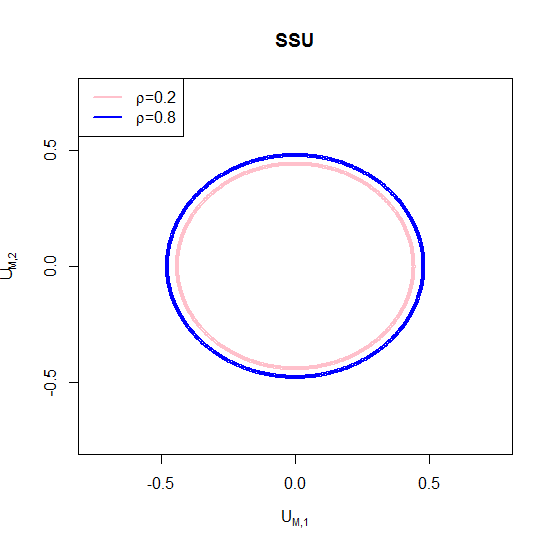}
%	\end{subfigure}
%	\begin{subfigure}[b]{0.3\textwidth}
	\includegraphics[width=3in]{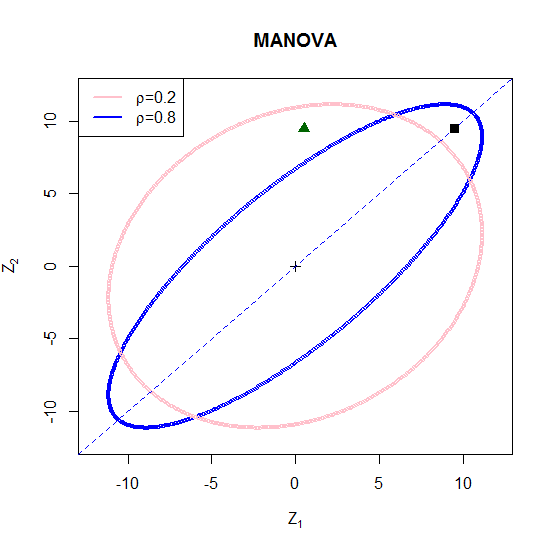}
%	\end{subfigure}
\caption[Comparison of SSU and MANOVA acceptance regions for $\rho=0.2,0.8$]{Comparison of theoretical $95\%$ acceptance regions of SSU test and of MANOVA for $K=2$ traits and $\rho=0.2,0.8$ (Compound Symmetry correlation). The area within the pink (blue) ellipse is the acceptance region when $\rho=0.2 \: (0.8)$. Details of these plots are provided in \ref{S-app2}.
SSU's acceptance region is drawn using the marginal scores $U_{M,1}$ and $U_{M,2}$. 
%A high true value of $\beta_1 (\beta_2)$ will be reflected by a high value of $U_{M,1} (U_{M,2})$. Irrespective of correlation, the acceptance regions are almost circular in shape. With increase in $\rho$, the shape of the acceptance region remains same and the direction of maximum variation for SSU does not change. Only the size increases a little which causes slight fall in power to reject $H_0$. \\
MANOVA's acceptance region is drawn based on the test $\bZ' \bI(\bs{0}) \bZ$, where $\bI(\bs{0})$ is Fisher Information matrix under $H_0: \bbeta=\bs{0}$. 
%A high true value of $\beta_1 (\beta_2)$ will be reflected by a high value of $Z_1 (Z_2)$.   
The blue dotted line is the $\bs{1}$ vector and coincides with the major axes of the ellipses. When $\beta_1=\beta_2$, we expect $Z_1 = Z_2$. The solid black square represents a situation where $\beta_1=\beta_2 \gg 0$.
% and MANOVA does not have power to reject $H_0$ at $5\%$ level when $\rho$ is high. 
When $\beta_1$ \& $\beta_2$ are significantly apart, we expect the same for $Z_1$ \& $Z_2$, and such a situation in represented by the dark green triangle.} %In this case, MANOVA is able to reject $H_0$ with greater power at high within trait correlation.}
 	\label{man-acc}
	\end{center}
\end{figure}

The theoretical $95\%$ acceptance regions of SSU and MANOVA for $K=2$ correlated traits in Figure \ref{man-acc} provide a geometrical explanation of the above theorems. The acceptance region of SSU is drawn using the marginal scores $U_{M,1}$ and $U_{M,2}$. 
MANOVA's acceptance region is drawn using the $2$ components $Z_1$ and $Z_2$ of vector $\bZ$ since MANOVA is asymptotically equivalent to the test $\bZ' \bI(\bs{0}) \bZ$. Here $\bZ$ is an $N(\bs{0}, \bI(\bs{0})^{-1})$ variable and $\bI(\bs{0})$ is Fisher Information matrix under $H_0: \bbeta=\bs{0}$. Details of this equivalence and the acceptance region plots are provided in \ref{S-app2}.
%[section \ref{app-manZ}].
For SSU, a high true value of $\beta_1 (\beta_2)$ will be reflected by a high value of $U_{M,1} (U_{M,2})$. 
In Figure \ref{man-acc}, observe that the SSU acceptance regions are almost circular in shape irrespective of correlation $\rho$. With increase in $\rho$, the shape of the acceptance region remains same.
% and the direction of maximum variation does not change. 
Only the size increases a little which causes slight loss in power to reject $H_0$.
For MANOVA, a high true value of $\beta_1 (\beta_2)$ will be reflected by a high value of $Z_1 (Z_2)$.   
%When $\beta_1=\beta_2$, we expect $Z_1 = Z_2$. The black square represents a situation where $\beta_1=\beta_2 \gg 0$
%and MANOVA does not have power to reject $H_0$ at $5\%$ level when $\rho$ is high. 
%When $\beta_1$ \& $\beta_2$ are significantly apart, we expect the same for $Z_1$ \& $Z_2$, and such a situation in represented by the dark green triangle. In this case, MANOVA is able to reject $H_0$ with greater power at high within trait correlation.
When $\rho \rightarrow 1$, notice that the acceptance region for MANOVA  becomes elongated along the direction of $\bs{1}$ vector in Figure \ref{man-acc}. Recall that for a CS correlation matrix, the eigen vector corresponding to the largest eigenvalue (for $\rho>0$) is along the direction of $\bs{1}$ vector. 
When the true genetic effect sizes are equal and in the same direction, the corresponding components of $\bZ$ are equal as well and they will lie on vector $\bs{1}$. 
%(the direction of maximal variability of the acceptance region of MANOVA). 
This suggests that the $Z$'s (and hence the non-zero genetic effects) need to be really large to cross the MANOVA acceptance region boundary for high $\rho$. The black box in Figure \ref{man-acc} represents such a situation, and
it arises when the SNP is associated with both the correlated traits.
This fail-to-reject situation will prevail even when the genetic effects are unequal but similar in magnitude. In genetic association studies, we may not expect equal effect sizes but we can expect them to be very close since each effect size is very small.
On the other hand, if the effect sizes are very different, the $\bZ$ vector will lie in some direction significantly away from the major axis $\bs{1}$ of the acceptance region. The closer it gets towards the minor axis, the greater is the  chance for MANOVA statistic to fall outside the boundary and reject the null. The dark green triangle in Figure \ref{man-acc} represents a situation where MANOVA's power to reject $H_0$ is higher when $\rho$ is higher. This is the situation when only one of the two traits is associated.
Furthermore, Figure \ref{man-acc} shows that MANOVA's loss in power will not be observed (irrespective of the strength and direction of within trait correlation) in studies where the effect sizes are reasonably large. This was observed in our simulation study with large genetic effects (simulation results not provided).
It is also to be noted that if all the traits are associated but not all are correlated, MANOVA is not expected to lose power 
(refer section \ref{sec:sim4}).

\subsection{An alternative test: A unified score-based association test (USAT)} \label{sec:usat}

Our proposed test is motivated by the geometrical findings in section \ref{sec:manprop}. As mentioned earlier, SSU test statistic does not explicitly incorporate within trait correlation and hence its acceptance region is not much affected when we increase the degree of dependency among the traits. On the other hand, MANOVA suffers from lack of power when the correlation is high and the genetic effect sizes are similar in magnitude and in same direction as the correlation. One, of course, does not know the true size and direction of the genetic effects and hence one would not know which association test to use. In such a scenario, one can see the clear advantage of combining MANOVA and SSU.
%We chose to weigh the test statistics from MANOVA and SSU, where 
We decided to choose the weight optimally from the data. We call our test unified score-based association test (USAT). The USAT test statistic is not exactly the best weighted combination of MANOVA and SSU. It is the minimum of the p-values of the different weighted combinations. \cite{Lee12} proposed a similar test statistic based on minimum p-value in the context of rare variants in sequencing association studies.
% since SSU is based on marginal scores while MANOVA is asymptotically equivalent to score test based on MMLR model \eqref{mmlr}.

Let $T_M$ be the MANOVA test statistic based on Wilk's lambda.
%, where $n$ is the number of unrelated individuals, $K$ is the number of traits, $q$ is the number of SNPs under consideration, and $\hat\lam_i$ is the sample $i$-th largest non-zero eigenvalue of $\hat\bSig_{11}^{-1/2} \hat\bSig_{12} \hat\Sig_{22}^{-1} \hat\bSig_{12}' \hat\bSig_{11}^{-1/2}$ (refer section \ref{sec:cca}). 
From Bartlett's approximation, $T_M \overset{a}\sim \chi^2_{K}$. 
On the other hand, the SSU test statistic, denoted as $T_S$, has an approximate $a\chi^2_d + b$ distribution, where the parameters $a$ and $b$ and the degrees of freedom $d$ are estimated from the data using equation \eqref{ssupar}. %So, $T_S := (T_{SSU} -b)/a \overset{a}\sim \chi^2_d$.
Consider the weighted statistic $T_\omg = \omg T_M + (1-\omg)T_S$, where $\omg \in [0,1]$ is the weight. Both MANOVA and SSU are special cases of the class of statistics $T_\omg$. Under the null hypothesis of no association, for a given $\omg$, $T_\omg$ is approximately a linear combination of chi-squared distributions. For a given $\omg$, the p-value $p_\omg$ of the test statistic $T_\omg$ can be calculated using \cite{Liu09-chi} algorithm for chi-square approximation of non-negative quadratic forms. It is worth noting that the calculation of $p_\omg$ does not require independence assumption of the two test statistics (refer \ref{S-app3}).

%The power of the test based on $T_\omg$ is maximized when $\omg$ is chosen appropriately. 
Apriori the optimal weight $\omg$ is not known. % and needs to be estimated from the data.
% by selecting the value of $\omg$ that maximizes power. 
%In other words, $T_{opt}=\{T_\omg : \min_{0\leq\omg\leq1} p_\omg \}$.
We propose our optimal unified test USAT as $$T_{USAT}=\min_{0\leq\omg\leq1} p_{\omg} $$
%In other words, the optimal p-value is $\min_{0\leq\omg\leq1} p_\omg$. 
For practical purposes, a grid of $11$ $\omg$ values were considered: $\{\omg_1=0,\omg_2=0.1,...,\omg_{10}=0.9,\omg_{11}=1\}$. A finer grid of more $\omg$ values did not change the USAT power curve much. 

To find the p-value of our USAT test statistic, we need the null distribution of USAT.
One option is to calculate the empirical p-value by considering several permuted datasets or by generating several datasets under the null (as done for Figure \ref{fig4}). Finding empirical p-values is computationally intensive and is not suitable when USAT is applied on a GWAS scale with large number of traits. We propose an approximate p-value calculation using a one-dimensional numerical integration. Observe that the p-value of statistic $T_{USAT}$ is
\begin{eqnarray*}
p_{USAT}&=&1-P(T_{USAT}\leq t_{USAT})\\
%&=& 1-P\lt(\min_\omg{p_\omg} \leq t_{USAT}\rt) = 1-P\lt(1-\min_\omg{p_\omg} > 1-t_{USAT}\rt) \\
%&=& 1-P\lt(\max_\omg{(1-p_\omg)} > 1-t_{USAT}\rt) \\
%&=& 1-P\lt( \{1-p_{\omg_1} > 1-t_{USAT}\}, \ldots, \{1-p_{\omg_{11}} > 1-t_{USAT}\}  \rt) \\
%&=& 1- P\Big( \{(1-p_{\omg_1})^{th} \text{ quantile} < (1-t_{USAT})^{th} \text{ quantile} \}, \ldots, \\
%& & \{(1-p_{\omg_{11}})^{th} \text{ quantile} < (1-t_{USAT})^{th} \text{ quantile} \}  \Big) \\
&=&1-P\lt(T_{\omg_1}<q_{\min}(\omg_1),..., T_{\omg_{11}}<q_{\min}(\omg_{11})\rt)
\approx 1 - \int F_{T_S}\big(\del_\omg(x)|x \big) f_{T_M}(x) dx
\end{eqnarray*}
where $t_{USAT}$ is the observed value of USAT test statistic for a given dataset,
 $q_{\min}(\omg_b)$ is the $(1-t_{USAT})$-th percentile of the distribution of $T_{\omg_b}$ for a given $\omg=\omg_b$, 
$F_{T_S}(.)$ is the cdf of SSU test statistic $T_S$, 
$\del_\omg(x)=\min_{\omg \in \{\omg_1,...,\omg_{11}\}} \frac{q_{\min}(\omg)-\omg x}{1-\omg}$
and  $f_{T_M}(.)$ is the pdf of MANOVA test statistic $T_M$. Mathematical details are provided in \ref{S-app3}.

\section{Results} \label{sec:result}

We compared the performances of different methods %in the three categories 
mentioned in sections \ref{sec:univ}, \ref{sec:ssu}, \ref{sec:multi}. We investigated their type I errors and powers by simulating data on unrelated subjects under a variety of trait models. 
%For our pilot studies, 
In Simulation 1 (section \ref{sec:sim1}), we considered $K=2$ correlated traits with genetic effects in different directions and correlation $\rho$ varying between $-1$ and $1$. For Simulation 2 (section \ref{sec:sim2}), we considered $K=5,10,20$ traits with genetic effects in the same direction as the positive correlation. CS correlation structure was considered.  As part of Simulation 2, we also compared the performance of USAT against MANOVA and SSU.
In Simulation 3 (section \ref{sec:sim3}), we used data from Simulation 2 and investigated the type I error of USAT using the p-value approximation method described in section \ref{sec:usat}. 
In Simulation 4 (section \ref{sec:sim4}), we used the same set-up as Simulation 2 to investigate the behaviors of existing methods under correlation structures other than CS.
%except the within trait correlation matrix. Correlation structures other than CS were used.
%In our simulation experiments, all the association tests except MultiPhen were coded by us in \verb+R 3.0.1+ \citep{R}. For MultiPhen, we used `Joint Model' output (p-value) from the \verb+R+ package \verb+MultiPhen 2.0.0+.

For our simulation studies, we first simulated $X$ taking values $0,1,2$ with probabilities $(1-f)^2, 2f(1-f), f^2$ respectively. $f=0.2$ was the m.a.f. of the the single SNP. The two alleles at the SNP were sampled independently to ensure Hardy-Weinberg equilibrium. Conditional on $X$, we simulated $\bY$ for a fixed $K$ using the simulation model
 $\bY = \beta_0\bs{1} + \bbeta X + \beps$, where the vectors $\bY$, $\bs{1}$, $\bbeta$, $\beps$ are $K$-dimensional. 
 We took $\beta_0=1$ and simulated $\beps$ from $N_K(\bs{0}, \sig^2\bR(\rho))$, where $\bR(\rho)$ is a CS correlation matrix. The specific choices of $\bbeta$, $\sig^2$ and $\rho$ for each simulation are given in the sections \ref{sec:sim1} and \ref{sec:sim2}.
% and $\sig^2$ was chosen such that the total variance of an associated trait is $10$. Possible values of $\rho$ considered were $-0.8,-0.6,-0.4,-0.2,0.2,0.4,0.6,0.8$. The choice of $\bbeta$ was based on fixed trait-specific QTL heritabilities, and our choices of heritabilities were not same for all the $K$'s we considered. With choice of $\bbeta$ varying with $K$, the total variance explained by the QTL also changed and hence our choice of $\sig^2$ in the residual covariance matrix differed for different values of $K$.
Before applying any method on the simulated datasets, we centered both $\bY$ and $\bX$ for each dataset.
We are interested in testing $H_0: \bbeta=\bs{0}$.
All the association tests except MultiPhen were coded by us in \verb+R 3.0.1+ \citep{R}. For MultiPhen, we used `Joint Model' output (p-value) from the \verb+R+ package \verb+MultiPhen 2.0.0+.

\subsection{Simulation 1: $K=2$ traits} \label{sec:sim1}
We first studied the performances of different association tests by considering only $2$ correlated traits so that the genetic effects and the pairwise correlation can have different directions. We considered genetic effects $\bbeta$ such that $0.2\%$ of the total variance of an associated trait was explained by the SNP. The total variance of an associated trait was taken to be $10$. This ensured that the variance due to SNP was $0.02$ while the residual variance was $\sig^2=9.98$. For an unassociated trait, the variance explained by SNP was $0$ and hence its residual variance was same as the total variance.
%The genetic effect of an unassociated trait is 0 (i.e., heritability is 0). 
We considered 3 possible levels of association: no trait was associated ($\beta_1=0=\beta_2$), only the first trait was associated ($\beta_1=0.25, \beta_2=0$) and both the traits were associated ($\beta_1=0.25=\beta_2$). We also considered genetic effects in opposite directions ($\beta_1=0.25, \beta_2=-0.25$). 
%8 values of $\rho$ were taken: $-0.8,-0.6,-0.4,-0.2,0.2,0.4,0.6,0.8$.
%We considered opposite signs of $\beta_1$ and $\beta_2$ to study the influence of a negative QTL-induced correlation.
% For each of $N=500$ datasets, $n=4,000$ individuals were simulated independently.

\begin{table}[b!]
\begin{center}
\caption[Type I error comparison of existing association tests]{Estimated type I errors of the afore mentioned existing association tests for $K=2$ correlated traits. $4$ values of pairwise correlation $\rho$ were considered. The p-values were calculated for $10,000$ null datasets with $4,000$ unrelated individuals. Type I error rate was calculated as the proportion of null datasets with p-value $\leq\alp$. }
	\label{t:type1k2}
\begin{tabular}{|cc|cccccc|}
\hline													
{$\alp$} & {$\rho$} & {Fisher} & {minP} & {TATES} & {SSU} & {MANOVA} & {MultiPhen} \\
%\hline
%$0.001$ & {$-0.8$} & $0.0065$ & $0.0009$ & $0.001$ & $0.0013$ & $0.0007$ & $0.0008$  \\
%$$ & {$-0.2$} & $0.0012$ & $0.0013$ & $0.0013$ & $0.0012$ & $0.0012$ & $0.0012$   \\
%$$ & {$0.2$} & $0.0014$ & $0.0013$ & $0.0013$ & $0.001$ & $0.0012$ & $0.0012$  \\
%$$ & {$0.8$} & $0.0066$ & $0.0007$ & $0.0009$ & $0.0009$ & $0.0006$ & $0.0008$  \\
\hline
$0.01$ & {$-0.8$} & $0.029$ & $0.009$ & $0.012$ & $0.010$ & $0.009$ & $0.010$  \\
$$ & {$-0.2$} & $0.012$ & $0.011$ & $0.011$ & $0.011$ & $0.011$ & $0.011$  \\
$$ & {$0.2$} & $0.011$ & $0.010$ & $0.010$ & $0.010$ & $0.011$ & $0.010$ \\
$$ & {$0.8$} & $0.026$ & $0.008$ & $0.012$ & $0.009$ & $0.009$ & $0.009$ \\
\hline
$0.05$ & {$-0.8$} & $0.079$ & $0.039$ & $0.053$ & $0.049$ & $0.049$ & $0.048$  \\
$$ & {$-0.2$} & $0.053$ & $0.051$ & $0.052$ & $0.049$ & $0.051$ & $0.051$  \\
$$ & {$0.2$} & $0.051$ & 0.049$$ & $0.051$ & $0.05$ & $0.052$ & $0.05$ \\
$$ & {$0.8$} & $0.079$ & $0.039$ & $0.055$ & $0.047$ & $0.052$ & $0.051$ \\
\hline
%$0.05$ & {$-0.8$} & $0.0788$ & $0.0397$ & $0.0529$ & $0.496$ & $0.0494$ & $0.0484$  \\
%$$ & {$-0.2$} & $0.0529$ & $0.0508$ & $0.0519$ & $0.495$ & $0.0508$ & $0.0511$  \\
%$$ & {$0.2$} & $0.0514$ & 0.0494$$ & $0.0512$ & $0.05$ & $0.0521$ & $0.05$ \\
%$$ & {$0.8$} & $0.0791$ & $0.0392$ & $0.0547$ & $0.0471$ & $0.0515$ & $0.0509$ \\
%\hline
\end{tabular}
\end{center}

\end{table}

First, type I error comparison was done for the 6 existing methods. For this purpose, we simulated $N=10,000$ null datasets with $n=4,000$ independent individuals. The type I error was calculated as the proportion of null datasets in which the p-value $\leq 0.01$ and $\leq0.05$. 
Table \ref{t:type1k2} shows the type I errors for each of the methods for $4$ values of $\rho$: $-0.8,-0.2,0.2,0.8$. For high magnitude of correlation $\rho$, notice that Fisher's method has inflated type I error while minP is conservative. Unlike minP, TATES is not conservative since it corrects for the relatedness among the traits. SSU maintains proper type I error since the distribution of the test statistic incorporates the within trait correlation structure. As expected, MANOVA and MultiPhen maintain correct type I error rate.

\begin{figure}[t!]
	\begin{center}
	\includegraphics[height=6in]{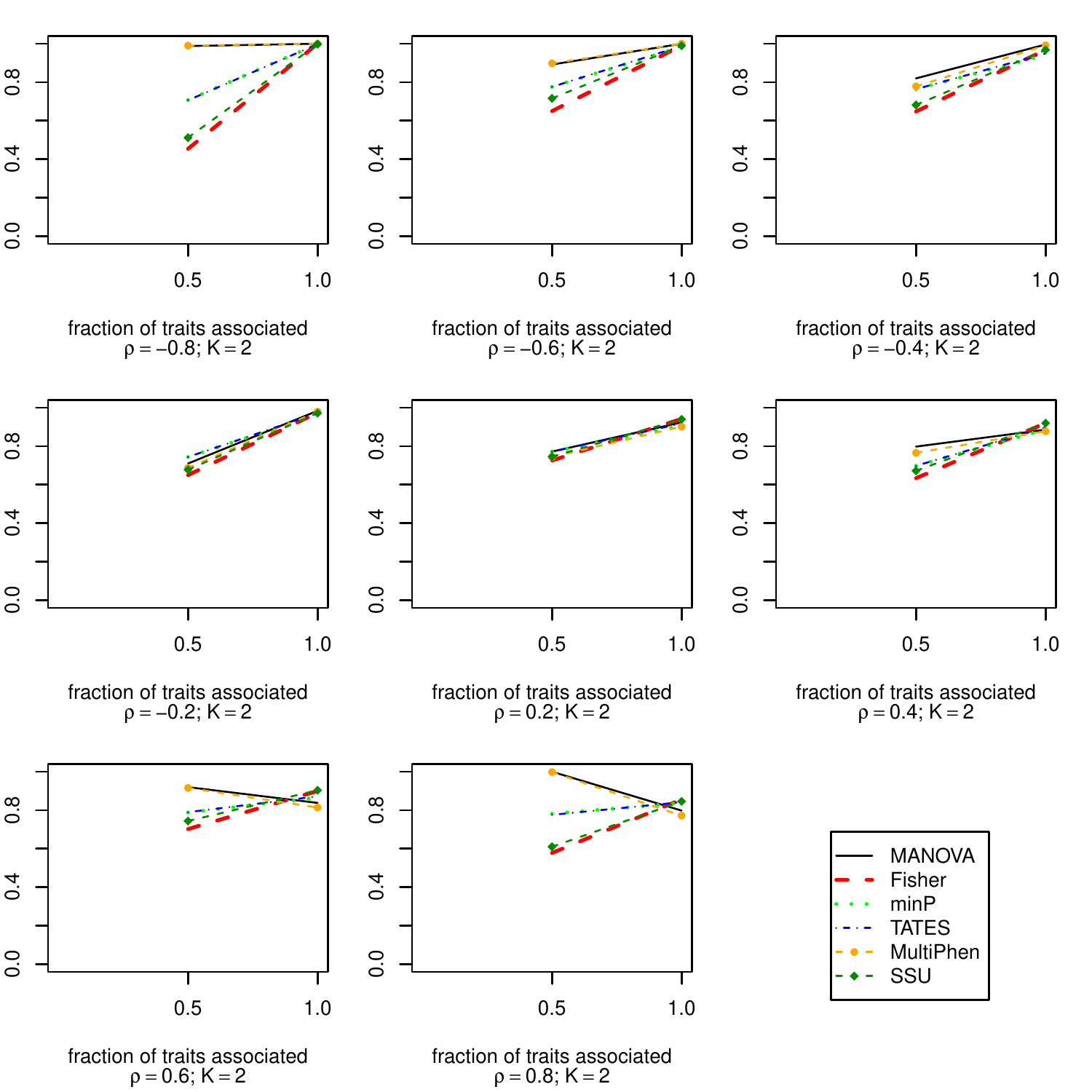}
	\caption[Simulation 1 power curves: 2 traits, same direction effects]{Empirical power curves of the different existing association tests for $K=2$ traits and different within trait correlation values $\rho=-0.8,-0.6,-0.4,-0.2,0.2,...,0.8$ based on $N=500$ datasets with $n=4,000$ unrelated subjects. Same direction and same size genetic effect used when both traits are associated (i.e., datasets are generated from an alternative model $H_{a2,1}:\beta_1=\beta_2>0$).  Effect size of $0.25$ is used for the associated traits. The power is plotted along y-axis while the fraction of traits associated with the genetic variant is plotted along x-axis.}
	\label{fig1}
	\end{center}
\end{figure}

Next, we compared the powers of the methods. $N=500$ datasets with $n=4,000$ unrelated individuals were simulated for different levels of association. $8$ different values of correlation $\rho$ were used: $-0.8,-0.6,-0.4,-0.2,0.2,0.4,0.6,0.8$. Since the methods do not have comparable type I errors (as seen in Table \ref{t:type1k2}), we plotted empirical power curves for comparison.
The empirical power at $5\%$ error level was calculated in the following way. 
For each of Fisher's method, MANOVA and SSU, the $95$-th quantile of the empirical distribution of the test statistic was determined based on the $N=500$ test statistics obtained from $N$ null datasets. Empirical power for these methods was calculated as the proportion of test statistics that exceeded the $95$-th quantile. 
For each of minP and TATES, the $5$-th quantile of the empirical distribution of the test statistic was determined using the $N=500$ test statistics under null. Empirical power was, then, calculated as the proportion of test statistics that could not exceed the $5$-th quantile. 
The empirical power of MultiPhen was determined using p-values in a way similar to empirical power calculation of minP and TATES. 
%$5\%$ critical values of the empirical distribution of the $N=500$ test statistics generated from the $500$ datasets where no trait was associated with the SNP.
From Figure \ref{fig1}, we observe that, irrespective of the value of $\rho$, the tests that do not consider within trait correlation have increase in power with increase in the number of associated traits. They seem to have similar performance when both traits are associated. 
%These methods, especially Fisher's test, do not maintain correct type I error, and this problem is more apparent at high values of $\rho$.  
On the other hand, MANOVA and MultiPhen have similar performance and are usually the most powerful approaches for detecting association. But, both experience power loss when $\rho>0.5$ and both traits have same direction of association. 
%Similar observations are made from Figure \ref{fig2}. 
For traits with genetic effects in opposite directions, similar behavior of MANOVA was observed %(refer Web Figure 1).
(refer Figure \ref{S-fig2} in \ref{S-app9}).
The power of MANOVA drops  when $\rho<-0.5$ and the 2 traits have opposite directions of association. 
These empirical observations on MANOVA are consistent with Corollary \ref{cor} of Theorem \ref{Thm2}.
No such power loss is observed for marginal model based approaches. In particular, SSU maintains correct type I error and does not experience power loss like MANOVA. This observation on SSU is consistent with our geometrical insight from Figure \ref{man-acc}.

%\begin{figure}[t!]
%	\begin{center}
%	\includegraphics[width=6in,height=6in]{./{trait2_opp_empirical}.pdf}
%	\caption[Simulation 1 power curves: 2 traits, opposite direction effects]{Empirical power curves of the different existing association tests for $K=2$ traits and different within trait correlation values $\rho=-0.8,-0.6,-0.4,-0.2,0.2,...,0.8$ based on $N=500$ datasets with $n=4,000$ unrelated subjects. Opposite direction but same size genetic effect used when both traits are associated (i.e., data is generated from an alternative model $H_{a2,2}:\beta_1=-\beta_2>0$). Effect size of $0.25$ is used for the associated traits. The power is plotted along y-axis while the fraction of traits associated is plotted along x-axis.}
%	\label{fig2}
%	\end{center}
%\end{figure}

%To study the effect of correlation $\rho$ and the directions of association on the performance of CCA, we plotted the empirical distribution of the test statistic $T_M$ under different possible alternatives.

\subsection{Simulation 2: $K=5,10,20$ traits} \label{sec:sim2}

To further study the performance of different tests with increase in the number of correlated traits, we simulated three sets of data where the first set had $K=5$, second had $K=10$ and the third had $K=20$ correlated traits. We considered $N=500$ simulated datasets for each scenario with $n=400$ unrelated individuals. For this simulation study, we considered only non-negative genetic effects, and positive correlation $\rho=0.2,0.4,0.6$ between each pair of traits. 
The total variance of a trait was fixed at $10$.
%We did not consider $\rho=0.8$ since it is unlikely for a large number of traits to have such high pairwise correlation. 
$\bbeta$ was chosen such that $0.5\%$ of the total variance of an associated trait was explained by the single SNP. 
% so that the residual variance is $\sig^2=9.95$. 
We considered 6 possible levels of association: $0\%$, $20\%$, $40\%$, $60\%$, $80\%$ or $100\%$ of the traits were associated with the SNP. Empirical power curves are presented for comparison.
%Here, empirical power curves were considered since the nominal type I errors of most methods differed from $0.05$ (as observed in Figures \ref{fig1} \& \ref{fig2}).  
%This ensured comparable type I error rate for all the methods.

\begin{figure}[t!]
	\begin{center}
	\includegraphics[height=7.5in]{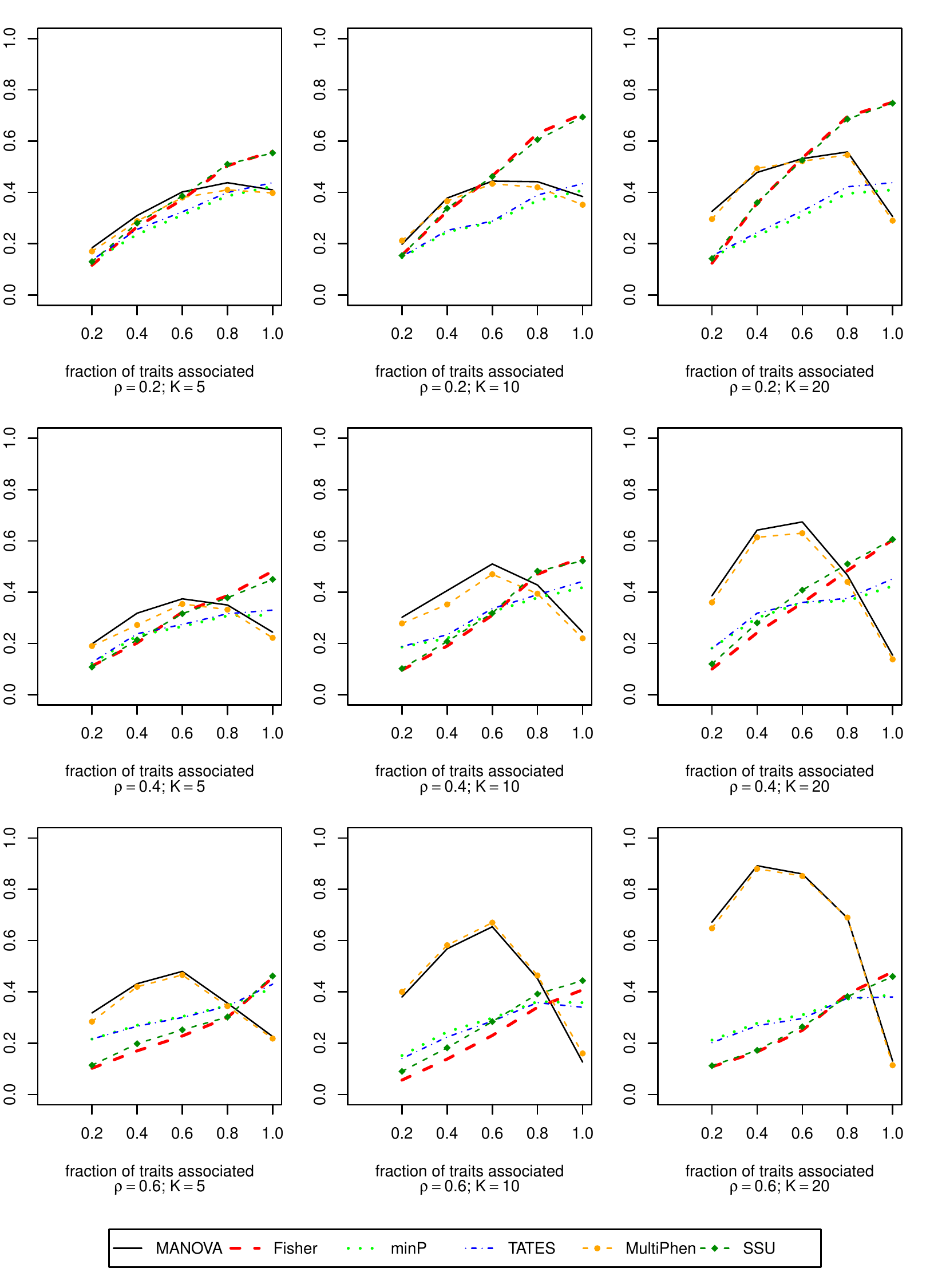}
	\caption[Simulation 2 power curves: 5,10,20 traits, same direction effects]{Empirical power curves of the different existing association tests for $K=5,10,20$ traits and different within trait correlation values $\rho=0.2,...,0.6$ based on $N=500$ datasets with $n=400$ unrelated subjects. 
%Nominal power curves were drawn for CCA/MANOVA, MultiPhen and TATES while empirical power curves were drawn for the rest. 
Same effect size of $0.395$ is used for all the associated traits. The power is plotted along y-axis while the fraction of traits associated with the genetic variant is plotted along x-axis.}
	\label{fig3}
	\end{center}
\end{figure}

From Figure \ref{fig3}, we again observe how MANOVA suffers from power loss at `complete association' when the within trait correlation is high. This power loss increases with increase in total number of correlated traits. 
At `complete association' (where MANOVA loses power), the power difference between MANOVA and other methods (such as SSU) increases with increase in number of correlated traits and decrease in correlation $\rho$. 
At a given `partial association', MANOVA is seen to dominate over other methods. Here, the difference in powers of MANOVA and any other method increases with increase in number of traits as well as the correlation. MANOVA's performance in this experiment is consistent with the asymptotic result in Theorem \ref{Thm3}.

\begin{figure}[t!]
	\begin{center}
	\includegraphics[height=7in]{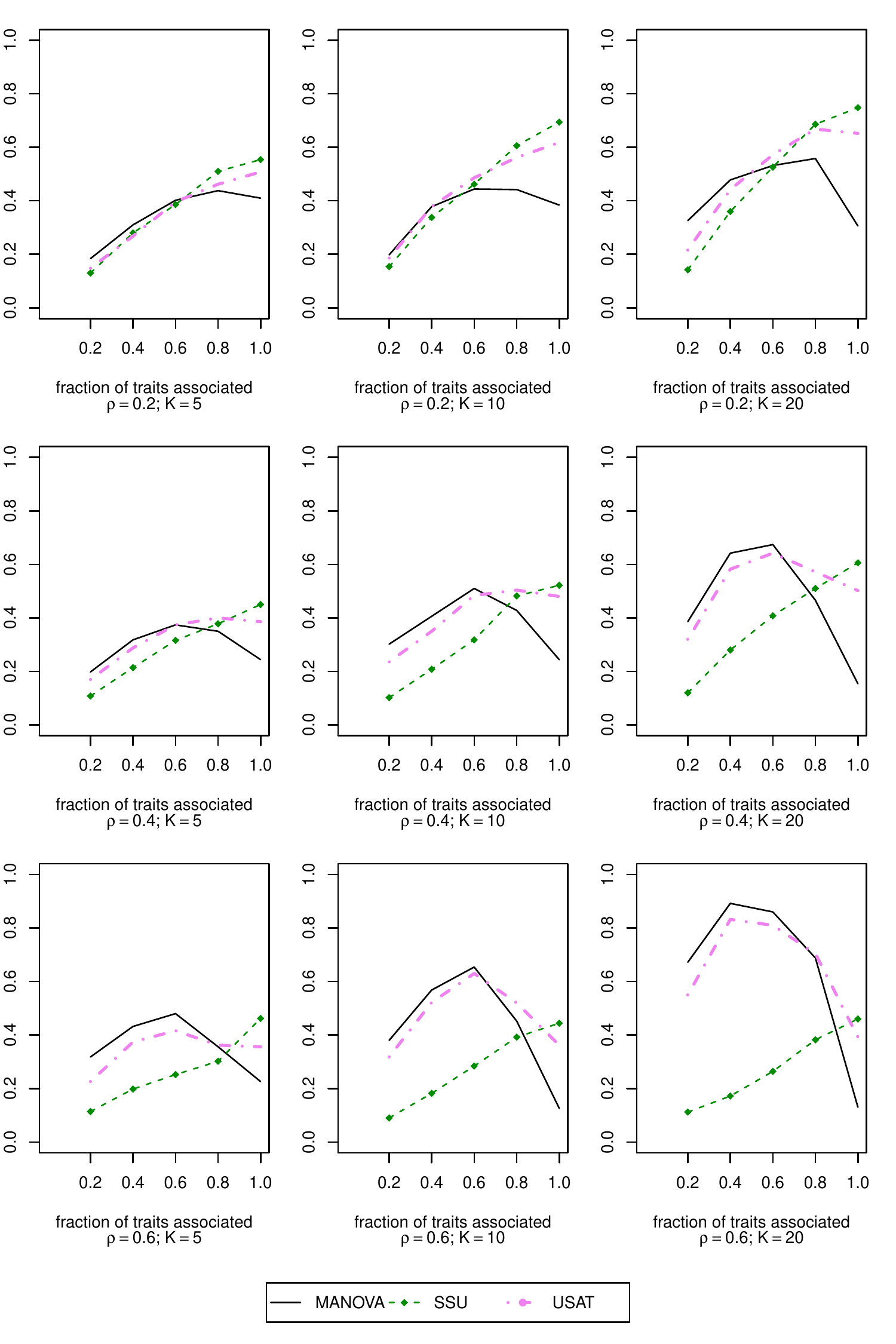}
	\caption[Comparison of USAT with SSU and MANOVA: 5,10,20 traits, same direction effects]{Empirical power curves of the SSU and MANOVA tests along with our novel approach USAT (based on an optimal combination of SSU and MANOVA). $K=5,10,20$ traits have been simulated at different within trait correlation values $\rho=0.2,0.4,0.6$. For each value of $K$ and $\rho$, there were $N=500$ datasets of $n=400$ unrelated individuals. Same effect size of $0.395$ was used for the traits that are associated. The power is plotted along y-axis while the fraction of traits associated with the genetic variant is plotted along x-axis.}
	\label{fig4}
	\end{center}
\end{figure}

Next we studied the performance of our approach USAT compared to MANOVA and SSU. 
%Since the sample size as well as the true genetic effects are small, the type I errors of the three approaches differed from $\alp=0.05$. So, empirical power curves were plotted in Figure \ref{fig4} for comparison. We did not consider very high trait correlation $\rho=0.8$ as it is unlikely to arise for large number of traits. 
We plotted empirical power curves in Figure \ref{fig4} for comparison.
Empirical powers for MANOVA and SSU were calculated as in section \ref{sec:sim1}.
Empirical power calculation of USAT was implemented in a way similar to minP and TATES (as described in section \ref{sec:sim1}).
%For USAT, the $5$-th quantile of the empirical distribution of the test statistic was determined using the $N=500$ test statistics from the $N$ null datasets. Empirical power was, then, calculated as the proportion of test statistics that could not exceed the $5$-th quantile.
Observe that USAT has better power than MANOVA whenever it suffers from power loss due to same direction of residual correlation and equal-sized genetic effects. In such situations, SSU performs significantly better than MANOVA, and USAT follows the SSU power curve closely.
%There is a brief geometric explanation in section \ref{sec:ssu} as to why SSU does better than MANOVA in such cases.
In other situations where MANOVA is seen to be most powerful among existing methods, USAT tends to have power close to MANOVA. USAT maximizes power by adaptively using the data to combine the MANOVA and the SSU approach.
% At high correlation and high number of traits, the behavior of our method approaches that of MANOVA.

\subsection{Simulation 3: p-value approximation for USAT} \label{sec:sim3}

In this section, we applied our approximate p-value approach for finding USAT p-values (discussed in section \ref{sec:usat}) to study its impact on type I error. We generated $N=100,000$ independent datasets (as in section \ref{sec:sim2}) with $n=10,000$ unrelated individuals under $H_0$.
%the null hypothesis of no association. 
The type I error was estimated by the proportion of datasets in which the asymptotic approximate p-value of USAT test statistic was $\leq 10^{-4}$, $\leq 10^{-3}$, $\leq 10^{-2}$, and $\leq 0.05$. 
Table \ref{t:type1} gives the estimated type I error rates for USAT using p-value approximation. The estimated values of type I error for different values of $K$ and $\rho$ were very close to the true error level $\alpha$.

%For the comparison of powers using empirical p-values and using the proposed p-value approximation for our SSU-CCA unified approach, . The single SNP was fixed to explain $0.05\%$ of the associated trait variances. For the empirical distribution of the unified test statistic, we generated $5,000$ null datasets under each scenario. The empirical power was then calculated as the fraction of datasets that had test statistic in the $100\alpha\%$ critical region of the empirical distribution.
%%In Figure \ref{fig5}, we see that the p-value approximation may give slightly inflated power for small number of traits $K$ at higher within trait correlation. For moderate to large values of $K$, the approximation method yields a power curve that is similar to the empirical power curve while being many times computationally faster than the empirical approach.

\begin{table}[t!]
%\begin{center}
%\caption{Estimated type I error of the approximate p-value calculation approach for our USAT test. The p-values were calculated for $10,000$ null datasets with $10,000$ unrelated individuals. Type I error rate is calculated as the proportion of datasets that had approximate p-value $\leq\alpha$. }
%	\label{t:type1}
%\begin{tabular}{|c||ccc||ccc||ccc|}
%\hline													
%{K} & {} & {5} & {} & {} & {10} & {} & {} & {20} & {} 	\\
%{$\rho$} & {0.2} & {0.4} & {0.6} & {0.2} & {0.4} & {0.6} & {0.2} &  {0.4} & {0.6} \\
%\hline
%{$\alpha=0.001$} & $0.0012$ & $0.0011$ & $0.0015$ & $0.0015$ & $0.0011$ & $0.0009$ & $0.0016$ & $0.0009$ & $0.001$ \\
%{$\alpha=0.01$} & $0.008$ & $0.010$ & $0.010$ & $0.009$ & $0.010$ & $0.009$ & $0.010$ & $0.011$ & $0.009$ \\
%{$\alpha=0.05$} & $0.033$ & $0.041$ & $0.046$ & $0.041$ & $0.047$ & $0.048$ & $0.044$ & $0.047$ & $0.046$ \\
%\hline
%\end{tabular}
%\end{center}

%\vspace{0.6cm}
\begin{center}
\caption[Type I error for USAT]{Estimated type I errors of the approximate p-value calculation approach for our USAT test. The p-values were calculated for $100,000$ null datasets with $10,000$ unrelated individuals. Type I error rate was calculated as the proportion of datasets that had approximate p-value $\leq\alpha$. }
{ \small
%\subfloat[First caption]{
	\label{t:type1}
%\begin{tabular}{|c||ccc||ccc||ccc|}
%\hline													
%{K} & {} & {5} & {} & {} & {10} & {} & {} & {20} & {} 	\\
%{$\rho$} & {0.2} & {0.4} & {0.6} & {0.2} & {0.4} & {0.6} & {0.2} &  {0.4} & {0.6} \\
%\hline
%{$\alpha=0.001$} & $0.0011$ & $0.0009$ & $0.0018$ & $0.0016$ & $0.0013$ & $0.0012$ & $0.0014$ & $0.0017$ & $0.0011$ \\
%{$\alpha=0.01$} & $0.0089$ & $0.0097$ & $0.0101$ & $0.0094$ & $0.0102$ & $0.0093$ & $0.0099$ & $0.0115$ & $0.0088$ \\
%{$\alpha=0.05$} & $0.0338$ & $0.0404$ & $0.0442$ & $0.04$ & $0.0435$ & $0.0449$ & $0.044$ & $0.0471$ & $0.0469$ \\
%\hline
%\end{tabular}
\begin{tabular}{|c||ccc||ccc||ccc|}
\hline													
{K} & {} & {5} & {} & {} & {10} & {} & {} & {20} & {} 	\\ \cline{2-4} \cline{5-7} \cline{8-10}
{$\rho$} & {0.2} & {0.4} & {0.6} & {0.2} & {0.4} & {0.6} & {0.2} &  {0.4} & {0.6} \\
\hline
{$\alpha=10^{-4}$} & $0.00008$ & $10^{-4}$ & $0.00008$ & $0.00015$ & $10^{-4}$ & $0.00008$ & $0.00015$ & $0.00011$ & $10^{-4}$ \\
{$\alpha=10^{-3}$} & $0.00081$ & $0.00092$ & $0.00086$ & $0.00108$ & $0.00099$ & $0.00089$ & $0.00149$ & $0.00117$ & $0.00109$ \\
{$\alpha=10^{-2}$} & $0.0082$ & $0.0091$ & $0.0093$ & $0.0094$ & $0.0098$ & $0.0097$ & $0.0103$ & $0.0104$ & $0.0103$ \\
{$\alpha=0.05$} & $0.0362$ & $0.0413$ & $0.0438$ & $0.0397$ & $0.0441$ & $0.0461$ & $0.0431$ & $0.0463$ & $0.0471$ \\
\hline
\end{tabular}
%}
}
\end{center}
\end{table}

\subsection{Simulation 4: Other correlation structures} \label{sec:sim4}

%Among the non-CS correlation structures that we considered, 
We first considered an independent structure. Apart from the residual correlation matrix $\bR(\rho)$, the data simulation was exactly same as in Simulation 2 (section \ref{sec:sim2}).
The figures and detailed explanations can be found in \ref{S-app4}. 
When all the traits are independent (i.e., $\bR(\rho)=\bI_K$), MANOVA does not suffer from power loss at any level of association. Empirical power curves (Figure \ref{S-fig4i}) showed that the performances of all the methods described in sections \ref{sec:univ}, \ref{sec:ssu}, \ref{sec:multi}, except minP and TATES, were similar. As expected, the powers steadily increased with increase in number of associated traits. 
Next we considered a correlation structure where the first $80\%$ of the traits had pairwise correlation $\rho$ while the rest were independent. %The correlation structure for the correlated traits was assumed to be CS. 
Empirical power curves (Figure \ref{S-fig4ii}) showed that MANOVA suffered power loss when only the correlated traits were associated. Performance of MANOVA improved when the SNP was associated with some of the uncorrelated traits. This simulation study
% with a bunch of correlated as well as uncorrelated traits 
showed us that MANOVA may not experience power loss even when all the traits are associated if some of them are uncorrelated. \ref{S-app5} provides a theoretical support for this observation. The third type of non-CS correlation structure that we considered was AR$1(\rho)$ (Figure \ref{S-fig4iii}). MANOVA's power loss was mainly observed for small $K$ and strong $\rho$. With increase in $K$ and decrease in $\rho$, MANOVA did not experience power loss even at `complete association'. 
%This is because 
The strength of AR$1(\rho)$ correlation becomes negligible at or near `complete association' when $\rho$ is small and $K$ is moderately large. 
For all these trait models, the power curves of marginal model based approaches rose with increase in number of associated traits (irrespective of strength or direction of residual correlation).
All these observations on MANOVA for various correlation structures were expected based on our geometrical insight from Figure \ref{man-acc} (section \ref{sec:manprop}).

\subsection{Real Data Analysis} \label{sec:real}

%Extensive evidence, including that gathered from twin and family studies, supports the hypothesis that genetic factors are a major contributor to the risk of type 2 diabetes (T2D).
The ARIC study is an ongoing prospective study designed to investigate the etiology and natural history of atherosclerosis and its clinical manifestations, and to measure variation in cardiovascular risk factors, medical care and disease by race, gender, place and time \citep{ARIC89}. ARIC has collected %fasting glucose measures along with other 
measures on many T2D-related traits at 4 separate visits over a 9-year period. 
For our analysis, we focused on the Caucasian participants and the following 3 T2D related quantitative traits %or risk factors 
measured at visit $4$ ($1996-98$): fasting glucose;  2-hour glucose from an oral glucose tolerance test; fasting insulin. %; BMI; waist circumference.  
The pairwise correlations among these 3 traits were within $(0.2,0.35)$. 
As in most studies of T2D, BMI was used as a covariate. 
Individuals with diagnosed or treated diabetes at visit $4$ and individuals with missing traits were excluded, leaving $5,816$ in our analytic sample. More details on the phenotypes and the choice of covariates can be found in \ref{S-app6}.

The ARIC cohort has been genotyped using the Affymetrix Genome-Wide SNP Array 6.0.
Genotyping was completed at the Broad Institute of MIT and Harvard in three batches; the
Birdseed algorithm was used for genotype calling. Imputation was performed using Mach
$1.0$ $86$ and HapMap release 21 (Build 35). SNPs with a call rate $< 90\%$, m.a.f. $< 1\%$, or
deviation from Hardy-Weinberg equilibrium ($p < 10^{-6}$) were excluded for imputation. 
There was a total of $2.5$ million genotyped or imputed SNPs.
%We performed a detailed analysis of the ARIC GWAS data using USAT and MANOVA to unravel novel SNPs %contributing to T2D risk.
%that possibly have influence on T2D.
Apart from USAT and MANOVA, we also performed separate univariate analyses to emphasize the importance of joint analysis over univariate ones. Before implementing any of these approaches, we centered both phenotype and genotype data. SNPs with m.a.f. $<1\%$ were excluded. All statistical models were adjusted for Age, Sex and BMI.

\begin{figure}[t!]
	\begin{center}
\vspace{-0.5cm}
	\includegraphics[width=6in,height=2in]{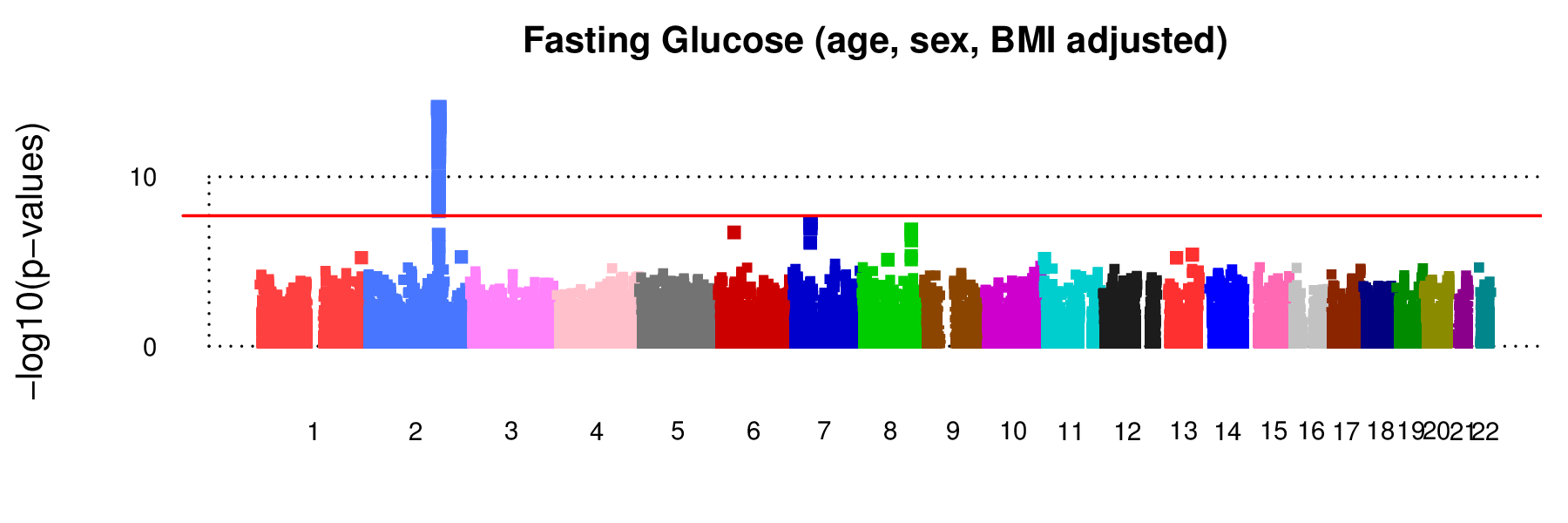} 		
	\includegraphics[width=6in,height=2in]{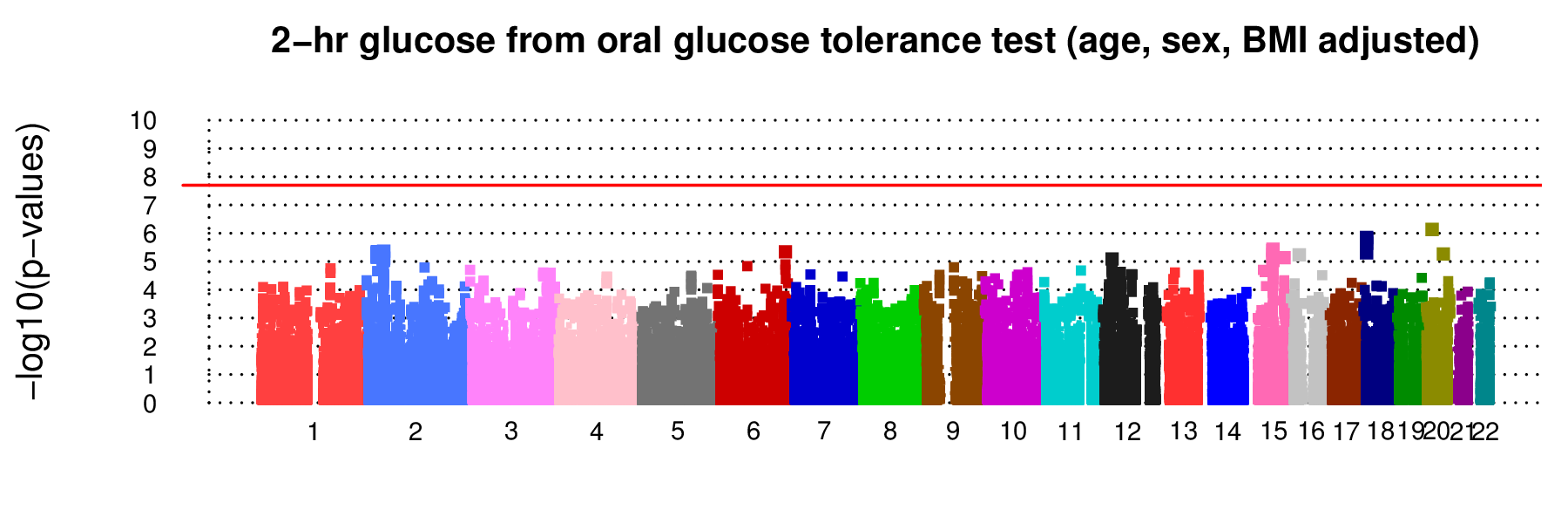}
	\includegraphics[width=6in,height=2in]{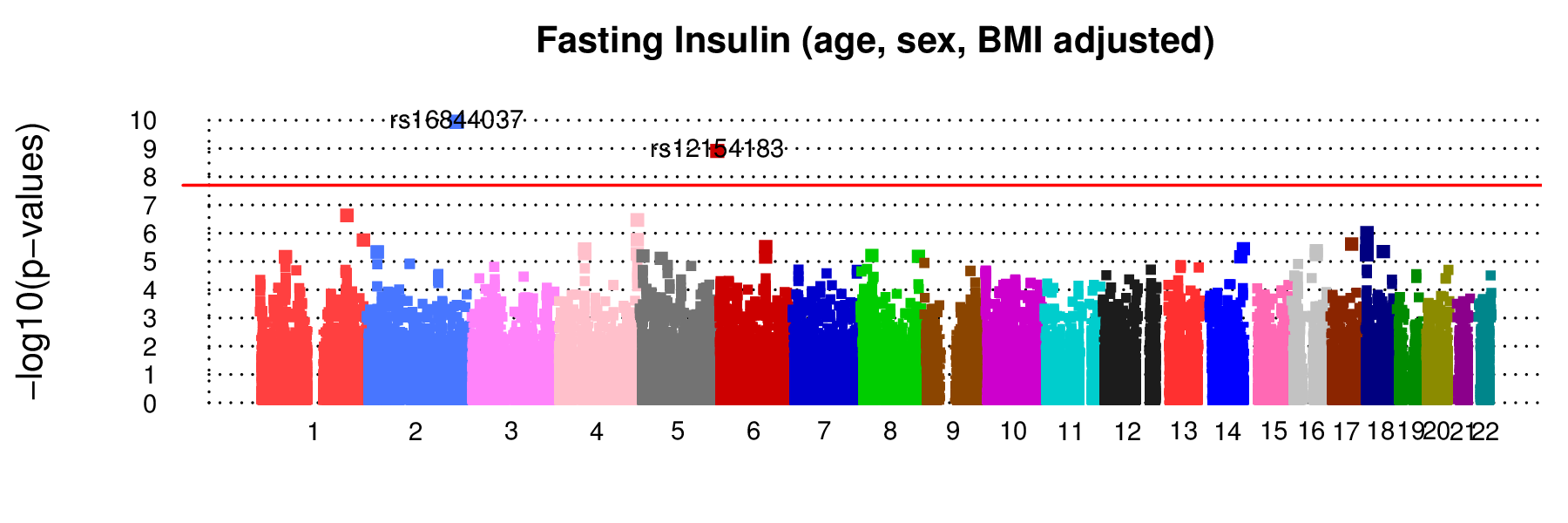}
	\caption[ARIC T2D GWAS: Manhattan Plots for Univariate Analyses]{ARIC Study: Manhattan plots of negative log-transformed p-values of the univariate analyses are plotted against base pair positions for chromosomes $1-22$.  Age, Sex and BMI were adjusted in the statistical models. The red horizontal line in each plot corresponds to significance level $0.05$ after Bonferroni correction for $2.5\times10^6$ SNPs. $53$ SNPs (all from chr 2) detected as significant for fasting glucose; $2$ such SNPs (from chr 2 and 6) for fasting insulin; and none for the other trait. Note that many of these significant SNPs are in high linkage disequilibrium.}
	\label{fig5}
	\end{center}
\end{figure}

\begin{figure}[b!]
	\begin{center}
\vspace{-0.5cm}
	\includegraphics[width=6in,height=4in]{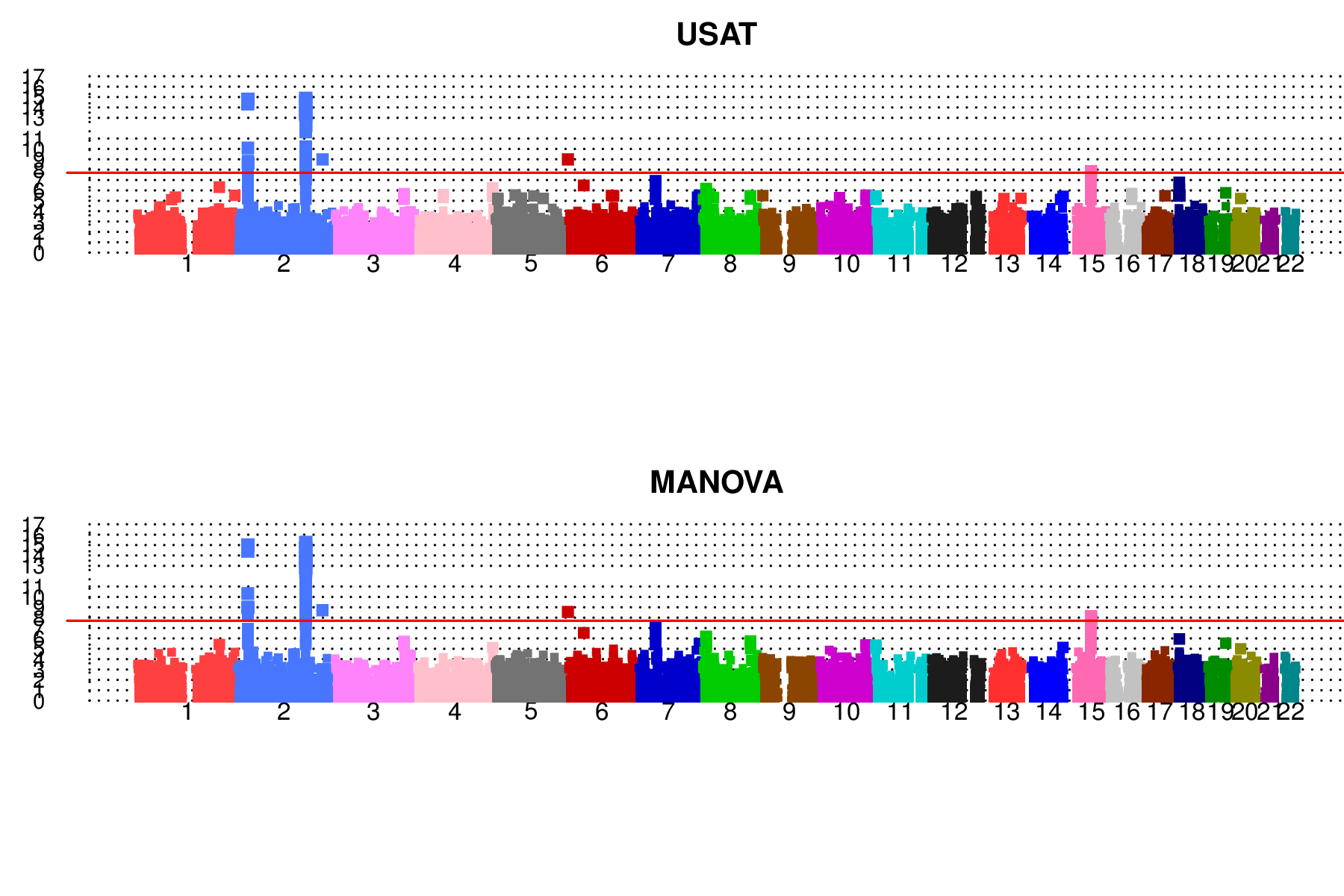}
	\caption[ARIC T2D GWAS: Manhattan Plots for MANOVA and USAT]{ARIC Study: Manhattan plots of negative log-transformed p-values of multivariate analyses (USAT and MANOVA) are plotted against base pair positions for chromosomes $1-22$. Age, Sex and BMI were adjusted in the statistical models. The red horizontal line in each plot corresponds to significance level $0.05$ after Bonferroni correction for $2.5\times10^6$ SNPs. 
Note that many of these significant SNPs are in high linkage disequilibrium.
}
	\label{fig6}
	\end{center}
\end{figure}

%We first conducted single trait single SNP association analysis.  
Figure \ref{fig5} shows the manhattan plots of negative log-transformed p-values for the single trait single SNP analyses for chromosomes $1-22$. The red horizontal line (at $7.7$) in each plot indicates the log-transformed GWAS significance p-value $2\times10^{-8}$. %$0.05$ after Bonferroni correction for the number of SNPs. 
There were $53$ and $2$ significant SNPs respectively for fasting glucose and fasting insulin. 
On the other hand, there were $86$ and $75$ signals for MANOVA and USAT respectively that reached this stringent Bonferroni corrected threshold (refer Figure \ref{fig6}, and %Web Table S1). 
Table \ref{S-t:sigall} in \ref{S-app8}).
Most of these signals mapped near the genes GCKR, ABCB11, C2orf16, CCDC121, ZNF512, FAM148A, C2CD4A, which are already known to be associated with diabetes related traits \citep[for example]{Yamauchi10, STAMPEED11}.
It is worth noting that these detected SNPs are in high linkage disequilibrium (LD). 
%For a complete list of these SNPs refer Table \ref{S-t:sigall}. %Web Table S1.
%The $11$ SNPs that USAT missed had p-values very close to MANOVA p-values. 
% were found to be significantly associated by both the multivariate methods.
Among the SNPs reported in %Web Table S1
Table \ref{S-t:sigall}, MANOVA and USAT respectively detected $36$ and $26$ SNPs that none of the univariate analyses could detect. Most notable genes that the univariate analyses completely missed are GCKR (on chr $2$) and FAM148A (on chr $15$).

%It is to be noted that many of the afore-mentioned associated SNPs are in high linkage disequilibrium (LD). 
Since most of the detected SNPs in Figure \ref{fig6} are in high LD, Table \ref{t:sig} reports only the important SNPs 
%(among the ones detected by MANOVA and USAT) 
after removing the ones in high LD.
In a group of highly correlated SNPs (i.e., SNPs with estimated absolute pairwise correlation coefficient $>0.8$ with another
SNP), we kept one SNP as a representative. 
The choice of representative SNP was based on previous reports of association.
The correlation coefficients (as measures of LD)
were obtained from PLINK \citep{Purcell07} using the command \texttt{plink --file mydata --r}. 
%Table \ref{t:sig} shows that there are 6 sets of independent signals spanning chr 2, one set in chr 6 and 2 such sets in chr 15.
The minor allele T of rs1260326 (gene GCKR of chr $2$) is known to be associated with T2D and hypertriglyceridemia. 
Risk allele A of rs13022873 (gene ZNF512 of chr $2$) was found to be significantly associated with waist circumference (a T2D related trait highly correlated with BMI) and triglycerides \citep{STAMPEED11}. 
rs13431652 (gene G6PC2 of chr $2$) was reported to be 
a potentially causative SNP linking G6PC2 to increased fasting plasma glucose levels and elevated promoter activity \citep{Diabetes10}.
%strongly associated with increased fasting plasma glucose level and elevated promoter activity, consistent with functional role of G6PC2 in pancreatic islets \citep{Diabetes10}.
The rs1402837 T allele (gene G6PC2 of chr $2$) is known to be associated with blood sugar levels (glycated hemoglobin levels).
\cite{McCaffery13} reported that SNPs in ABCB11 (like rs484066) of chr $2$  are associated with weight loss and regain. 
Meta-analysis of several GWAS found rs17271305 (gene VPS13C of chr $15$) to be associated with glucose levels 2 hours after an oral glucose challenge \citep{Saxena10}.
The diabetogenic A allele of rs7172432 (gene VPS13C/C2CD4A/C2CD4B of chr $15$) significantly impairs glucose-stimulated insulin response in non-diabetics \citep{Grarup11}.
The remaining signals in Table \ref{t:sig} have not been previously reported.

\begin{table}[t!]
\begin{center}
\caption[ARIC T2D GWAS: Significant SNPs from MANOVA and USAT]{List of SNPs that exceed the significance level of $0.05$ after Bonferroni correction for the number of SNPs (i.e., p-value threshold $2\times10^{-8}$) for USAT and MANOVA. The SNPs listed here are the ones left after LD screening. In a group of highly correlated SNPs (i.e., SNPs with estimated absolute pairwise correlation coefficient $>0.8$ with another SNP), one SNP was kept as a representative. $p$ values for the univariate analysis of the individual traits are also provided for these significant SNPs. SNP rs ID in bold is the one detected solely by MANOVA but not by USAT at this stringent genome-wide significance level. The abbreviations used are FG (Fasting Glucose), 2-hr GL (2-hour glucose from an oral glucose tolerance test), FI (Fasting Insulin).}
	\label{t:sig}

{ \small
\begin{tabular}{|c|c|c|c|c|c|c|c|}
\hline												
  &  &  & MANOVA & USAT &  \multicolumn{3}{c}{Univariate Analysis $p$}\\
  chr & SNP & position & $p$ & $p$ & FG & 2-hr GL & FI \\ \hline

$2$ & $rs1260326$ & $27584444$ & $3.77\times10^{-15}$ & $4.44\times10^{-15}$ & $1.24\times10^{-4}$ & $6.26\times10^{-6}$ & $1.24\times10^{-5}$ \\

%$2$ & $rs4665987$ & $27609329$ & $9.67\times10^{-10}$ & $2.31\times10^{-9}$ & $1.56\times10^{-2}$ & $4.49\times10^{-6}$ & $1.46\times10^{-2}$ \\
$2$ & $rs13022873$ & $27669014$ & $9.94\times10^{-10}$ & $2.34\times10^{-9}$ & $1.49\times10^{-2}$ & $6.01\times10^{-6}$ & $1.07\times10^{-2}$ \\

$2$ & $rs13431652$ & $169461661$ & $1.85\times10^{-13}$ & $5.48\times10^{-13}$ & $2.24\times10^{-12}$ & $9.57\times10^{-1}$ & $2.85\times10^{-1}$ \\

$2$ & $rs1402837$ & $169465600$ & $4.91\times10^{-9}$ & $1.15\times10^{-8}$ & $2.78\times10^{-10}$ & $5.18\times10^{-2}$ & $8.07\times10^{-1}$ \\

%$2$ & $rs478333$ & $169487402$ & $8.61\times10^{-10}$ & $1.42\times10^{-9}$ & $3.33\times10^{-10}$ & $4.16\times10^{-1}$ & $6.41\times10^{-1}$ \\
$2$ & $rs484066$ & $169490727$ & $2.01\times10^{-12}$ & $2.32\times10^{-12}$ & $6.10\times10^{-12}$ & $8.54\times10^{-1}$ & $4.87\times10^{-1}$ \\

$2$ & $rs16844037$ & $210561151$ & $1.92\times10^{-9}$ & $1.04\times10^{-9}$ & $2.16\times10^{-1}$ & $2.56\times10^{-2}$ & $1.14\times10^{-10}$ \\

$6$ & $rs12154183$ & $295724$ & $2.81\times10^{-9}$ & $1.02\times10^{-9}$ & $4.01\times10^{-1}$ & $9.55\times10^{-1}$ & $1.21\times10^{-9}$ \\

%$15$ & $\bs{rs7170293}$ & $60023665$ & $1.98\times10^{-8}$ & $6.28\times10^{-8}$ & $5.77\times10^{-3}$ & $3.20\times10^{-5}$ & $2.68\times10^{-1}$ \\
$15$ & $rs17271305$ & $60120272$ & $6.87\times10^{-9}$ & $1.35\times10^{-8}$ & $6.29\times10^{-3}$ & $1.17\times10^{-5}$ & $2.86\times10^{-1}$ \\

%$15$ & $\bs{rs7163757}$ & $60178900$ & $1.68\times10^{-8}$ & $5.98\times10^{-8}$ & $7.98\times10^{-4}$ & $3.52\times10^{-4}$ & $4.88\times10^{-2}$ \\
$15$ & $\bs{rs7172432}$ & $60183681$ & $1.68\times10^{-8}$ & $5.98\times10^{-8}$ & $7.98\times10^{-4}$ & $3.52\times10^{-4}$ & $4.88\times10^{-2}$ \\
\hline
\end{tabular}
}
\end{center}
\end{table}

In Table \ref{t:sig}, we notice one SNP (rs7172432) that USAT missed at the stringent significance level of $2\times10^{-8}$. One also notices that difference in the p-values of USAT and MANOVA for this SNP is negligible.
If one takes a closer look at the manhattan plots of Figure \ref{fig6}, one will find that certain SNPs are prominently visible for USAT but not for MANOVA  (even though none could reach genome wide significance). The most noticeable regions are in 
chr $1$ (rs4427409 and rs17434403 with USAT $p$ $4.95\times10^{-7}$ and $4.86\times10^{-7}$), 
chr $4$ (rs4861722 and rs11729070 with USAT $p$ $5.76\times10^{-7}$ and $3.86\times10^{-7}$) and in 
chr $18$ (rs17497377 and rs2864527 with USAT $p$ $1.73\times10^{-7}$ and $7.59\times10^{-7}$). 
None of these signals have been previously reported.

\section{Discussion} \label{sec:discuss}
In the study of a complex disease, several correlated traits are often measured as risk factors for the disease. There may be genetic variants affecting several of these traits. Analyzing multiple disease-related traits could potentially
increase power to detect association of genetic variants with such a disease. The elucidation of genetic risk factors of such diseases will help us in better understanding and developing therapeutics against them.
In this paper, we have studied some of the common univariate and multivariate approaches for analyzing association between multiple phenotypes and a genetic variant. 
Our simulation results showed that no single method perform uniformly better than the others under the simulation scenarios we considered. Multivariate methods like MANOVA and MultiPhen usually had higher power than the univariate tests only in situations where a few of the correlated traits were associated. 
%Multivariate methods also performed better when the direction of genetic effects was not same as the direction of dependence among the traits. 
Univariate model based methods in sections \ref{sec:univ} \& \ref{sec:ssu} outperformed multivariate methods when all the correlated traits were associated and the genetic effects as well as the residual correlations were in the same direction. 
%Most of the methods in our simulation study have been compared to one another  by references stated earlier but none of them provided any theoretical justification. 
Under the assumption of a CS residual correlation structure, we established theoretical conditions 
%(involving genetic effects and residual correlation) 
as to when MANOVA  would start losing power. Although we have not established similar theoretical conditions for other correlation structures, we have seen similar behavior of MANOVA in our simulation studies. %(section \ref{sec:sim4}). 

We also proposed a novel weighted approach USAT, which maximizes power by adaptively using the data to optimally combine MANOVA and the SSU test. 
%For USAT, one can either compute empirical p-values to maintain correct type I error rate, or compute approximate p-values using a very fast one-dimensional numerical integration. 
%Calculation of approximate p-value is recommended when this approach is applied on GWAS data. 
Approximate USAT p-values can be computed using a very fast one-dimensional numerical integration, which makes implementation on GWAS data easy.
As shown by our simulation studies, USAT maintains correct Type I error (refer Table \ref{t:type1}) and has good power in detecting association (refer Figure \ref{fig4}). Unlike MANOVA, USAT is powerful in detecting pleiotropy under the simulation models we considered. 
The ARIC data analysis not only emphasized the importance of joint analysis of correlated endo-phenotypes over univariate analyses but also showed the power of USAT in detecting SNPs that might have influence on T2D risk.
As in the real data analysis, adjustment of other covariates can be easily done for USAT (details in \ref{S-app7}). % for details).

Finally, the simulation scenarios we considered are not exhaustive. Under the scenarios we considered, we found it best to combine the SSU and the MANOVA tests. The relative behavior of these two tests did not vary much with change in m.a.f. %(refer Web Figure 2)
(refer Figure \ref{S-fig4-maf5} in \ref{S-app10}), or with increase in the number of correlated traits.
%, or with change in correlation structure from CS to AR1 to a structure where some were correlated and some were not. 
Our simulation studies also assumed no missing data and no trait outliers. USAT requires complete phenotype data. In presence of missing traits, one may consider imputation before performing association analysis. \cite{Sluis13} showed that $10\%$ missing-completely-at-random data caused quite a drop in power for MANOVA when only 1 trait was associated. \cite{OReilly12} showed that in the presence of outliers in the phenotype distribution, MANOVA and the standard univariate approach were substantially inflated for low m.a.f.
We simulated data for an additive model only and did not consider any non-additive genetic model and/or interactions.
%, which may be more likely in real data. 
In future, we intend to study how power of %SSU test, and hence 
our USAT test would be affected in such situations. 
%One must also note that our version of the SSU test assumed a binomial categorization of the genetic variant while MANOVA considered all 3 possible values. Empirically, we found no significant loss in power for such a binomial assumption for m.a.f. $0.2$ or lower (refer Appendix S5). Instead of the logistic regression, perhaps a more appropriate approach is to use a proportional odds regression for SSU which will allow us to use all possible values the variable $X$ can take. However, we found this approach unnecessarily complicated to implement, and hence went ahead with binomial categorization for the SSU test.

%\subsection{Real Data Analysis} \label{sec:real}

				%%%%%%%%%%%%%%%%%%%%%%%%%
\backmatter

\section*{Acknowledgments}
This research was supported by NIH grant R01-DA033958 (PI: Saonli Basu), the Doctoral Dissertation Fellowship of the University of Minnesota Graduate School and the Minnesota Supercomputing Institute. 
%We are grateful to the Minnesota Supercomputing Institute (MSI) for partial support of this work.
The ARIC Study is carried out as
a collaborative study supported by National Heart, Lung, and Blood Institute contracts
(HHSN268201100005C, HHSN268201100006C, HHSN268201100007C,
HHSN268201100008C, HHSN268201100009C, HHSN268201100010C,
 HHSN268201100011C, HHSN268201100012C), R01HL087641, R01HL59367 and
R01HL086694;
National Human Genome Research Institute contract U01HG004402; 
and NIH contract 
HHSN268200625226C.
Infrastructure was partly supported by Grant Number UL1RR025005, a component of the NIH and NIH Roadmap for Medical Research.
We thank the staff and participants of the ARIC study for their important contributions.
The authors have no conflict of interests to declare.

\section*{Supporting Information}
Appendices S1$-$S10
are available with this paper at the end.

\bibliographystyle{biom}	% (uses file "plain.bst")
\bibliography{oralprop}

\label{lastpage}
\end{document}

\begin{center}
%\section*{Web-based Supplementary Materials}
{\LARGE\bfseries{Supporting Information}}\\
\sc{for} \\
\textbf{USAT: A Unified Score-based Association Test for 
%Population-based 
Multiple Phenotype-Genotype Analysis}\\
\sc{by}\\

\footnotesize{Debashree Ray$^1$, James S. Pankow$^2$, Saonli Basu$^1$} \\
\footnotesize{$^1$\emph{Division of Biostatistics, School of Public Health, University of Minnesota, U.S.A.}} \\
\footnotesize{$^2$\emph{Division of Epidemiology \& Community Health, School of Public Health, University of Minnesota, U.S.A.}}
\end{center}

\vspace{1cm}
%%%%%%%%%  Web Appendix A
\subsection{}		\label{app1}

%\no\textbf
\subsubsection{Proof of Theorem~\ref{M-Thm3} } 

Without loss of generality, we assume that $\bY$ and $\bX$ are centered. For testing $H_0: \bbeta=\bs{0}$, the Wilk's Lambda test statistic is $\text{det}\bE/\text{det}(\bH+\bE)=\text{det}(\frac{1}{n}\bE)/\text{det}(\frac{1}{n}\bH+\frac{1}{n}\bE)$ , where $\bH = \hat\bbeta (\bX'\bX) \hat\bbeta'$, $\bE = \bY'\bY - \hat\bbeta (\bX'\bX) \hat\bbeta'$, $n$ is the number of unrelated individuals, and $\hat\bbeta = \bY' \bX (\bX'\bX)^{-1}$ is the least squares estimate of the vector of genetic effects $\bbeta$. Note that $\bX'\bX = \sum_{i=1}^n X_i^2$ is a random variable (not a matrix), where $\E(X_i^2)=2f(1-f)=\Var(X_i) \: \forall \: i$.
 Using our distributional assumptions about centered $\bX$ and $\Eps$, it can be shown that 
$ \frac{1}{n}\bH \overset{\P}\rightarrow 2f(1-f) \bbeta \bbeta'$ and
$\frac{1}{n}\bE \overset{\P}\rightarrow \bSig$ as $n\rightarrow\infty$. 
Here, $\overset{\P}\rightarrow$ denotes convergence in probability as $n\rightarrow \infty$.

For the CS residual covariance matrix $\bSig$, we know that the eigen vector corresponding to the largest eigenvalue $\lam_{1}=\sig^2\{1+(K-1)\rho\}$ is $\bv_1 \propto \bs{1}$, while the eigen vectors corresponding to $\lam_{2}=...=\lam_{K}=\sig^2(1-\rho)$ are respectively $\bv_2,...,\bv_K$ such that $\bs{1}'\bv_k=0 \: \forall \: k=2,...,K$. For the eigen vectors to be orthonormal, we must have $\bv_1=c_K \bs{1}$ where $c^2_K=1/K$.
Thus, we can write,
$\bSig =\lam_{1} c^2_K \bs{1}\bs{1}' + \sum_{i=2}^K \lam_{i} \bv_i \bv'_i$ and
$\bSig^{-1} = \frac{1}{\lam_{1}} c^2_K \bs{1}\bs{1}' + \sum_{i=2}^K \frac{1}{\lam_{i}} \bv_i \bv'_i$.

Consider the testing of $H_0: \bbeta=0$ against two possible alternatives: $H_{a,u}: \beta_1=...=\beta_u\neq0, \beta_{K-u}=...=\beta_K=0$ (partial association) and $H_{a,K}: \beta_1=...=\beta_K\neq0$ (complete association). 
%Here, for the partial association case, $0<u<K$ is the number of traits associated and $u/K$ is the proportion of associated traits among the correlated traits. 
Under the alternative $H_{a,K}$ (complete association), $|\bI + \bH\bE^{-1}|$ is given by
\begin{eqnarray*}
%|\bI + \bH\bE^{-1}| \overset{H_{a,K}}{:=} 
\lt|\bI + \frac{\bH_K}{n}\lt(\frac{\bE}{n}\rt)^{-1}\rt| 
&\overset{P}{\underset{n\rightarrow \infty}\rightarrow}&
 \lt| \bI_K + (2f(1-f) \beta_1^2 \bs{1}\bs{1}') \lt(\frac{1}{\lam_1} c^2_K \bs{1}\bs{1}' + \sum_{i=2}^K \frac{1}{\lam_i} \bv_i \bv'_i\rt) \rt| \\
&=&1+\frac{2f(1-f)\beta_1^2}{\lam_1} K
\end{eqnarray*}
%\big(using the fact that $|\bA+\bx\bx'| = |\bA| (1+\bx'\bA^{-1}\bx)$, provided $|\bA|\neq0$, and hence 
%$|a\bI_K + b \bs{1}\bs{1}'| = a^K (1+\frac{b}{a}K)$ \big)
Under the alternative $H_{a,u}$ (partial association),  

$|\bI + \bH\bE^{-1}| \overset{H_{a,u}}= |\bI + \frac{\bH_u}{n}\lt(\frac{\bE}{n}\rt)^{-1}| $
\begin{eqnarray*}
% |\bI &+& \bH_u\bE^{-1}| \\
%&\overset{P} {\underset{n\rightarrow \infty}\rightarrow}& 
%|\bI + \bH\bE^{-1}| &\overset{H_{a,u}}=& |\bI + \frac{\bH_u}{n}\lt(\frac{\bE}{n}\rt)^{-1}| \\
&\overset{P} {\underset{n\rightarrow \infty}\rightarrow}&\lt| \bI_K + 2f(1-f) \lt( \begin{matrix} 
\beta_1^2 \bs{1}_{u}\bs{1}'_{u} & \bs{0} \\ \bs{0}' & \bs{O} 
\end{matrix} \rt)
\lt( \begin{matrix} \bSig_{11(u\times u)} & \bSig_{12(u\times \overline{K-u})} \\ \bSig'_{12(\overline{K-u}\times u)} & 
\bSig_{22(\overline{K-u} \times \overline{K-u})} \end{matrix} \rt)^{-1} \rt| \\
&=& \lt| \bI + 2f(1-f) \lt( \begin{matrix} 
\beta_1^2 \bs{1}_u\bs{1}'_u & \bs{0} \\ \bs{0}' & \bs{O} 
\end{matrix} \rt)
\lt( \begin{matrix} 
\bSig^{11} & \bSig^{12} \\ 
\star &  \star \end{matrix} \rt) \rt| \\
&=& \lt| \bI + 2f(1-f) \lt( \begin{matrix} \bA & \bB \\ \bs{0}' & \bs{O} \end{matrix} \rt) \rt| \\
&=& \lt| \bI_u + 2f(1-f) \bA  \rt| \\
% &=& \lt| \bI_u + \frac{2f(1-f)\beta_1^2}{\sig^2(1-\rho)}\times\frac{1+(K-u-1)\rho}{1+(K-1)\rho} \bs{1}_u \bs{1}'_u \rt| \\
&=& 1+ \frac{2f(1-f)\beta_1^2}{\sig^2(1-\rho)}\frac{1+(K-u-1)\rho}{1+(K-1)\rho} u
\end{eqnarray*}
where $\bSig_{11} = \sig^2 (1-\rho) \bI_u + \sig^2 \rho \bs{1}_u\bs{1}_u'$, 
$\bSig_{22}=\sig^2 (1-\rho) \bI_{K-u} + \sig^2 \rho \bs{1}_{K-u}\bs{1}_{K-u}'$, 
$\bSig_{12}=\sig^2\rho \bs{1}_u\bs{1}'_{K-u} $,
$\bSig^{11} = ( \bSig_{11} - \bSig_{12}\bSig_{22}^{-1} \bSig'_{12} )^{-1}$, 
$\bSig^{12} = -\bSig_{11}^{-1} \bSig_{12}(\bSig_{22} - \bSig'_{12} \bSig_{11}^{-1} \bSig_{12})^{-1}$, \\
$\bA = \beta_1^2 \bs{1}_u \bs{1}_u' \lt( \bSig_{11} - \bSig_{12}\bSig_{22}^{-1} \bSig'_{12} \rt)^{-1}$,
$\bB = -\beta_1^2 \bs{1}_u \bs{1}_u' \bSig_{11}^{-1} \bSig_{12}(\bSig_{22} - \bSig'_{12} \bSig_{11}^{-1} \bSig_{12})^{-1}  $

%It is now straightforward to show that
%\begin{eqnarray*}
%\bSig_{11} - \bSig_{12}\Sig_{22}^{-1} \bSig'_{12}
%&=& \sig^2(1-\rho) \bI_u + \sig^2\rho \bs{1}_u\bs{1}_u' - \sig^2 \rho^2 \frac{K-u}{1+\rho(K-u-1)} \bs{1}_u\bs{1}'_{u} \\
%&=& a\bI_u + b\bs{1}_u\bs{1}'_u, \text{ where }a= \sig^2(1-\rho), \: b= \sig^2\rho \frac{(1 - \rho)}{1+\rho(K-u-1)} \\
%%
%\therefore \: \lt( \bSig_{11} - \bSig_{12}\Sig_{22}^{-1} \bSig'_{12} \rt)^{-1}
%&=& \frac{1}{a} \bI_u - \frac{b}{a(a+ub)}\bs{1}_u\bs{1}'_u 
%= \frac{1}{\sig^2(1-\rho)} \lt[ \bI_u - \frac{\rho}{1+(K-1)\rho}\bs{1}_u\bs{1}'_u \rt] \\
%%
%\therefore \bA 
%&=&  \frac{\beta_1^2}{\sig^2(1-\rho)} \lt[ 1 - \frac{u\rho}{1+(K-1)\rho} \rt] \bs{1}_u\bs{1}'_u \\
%%
%\therefore |\bI_u + 2f(1-f) \bA|
%% &=& \lt| \bI_u + \frac{2f(1-f)\beta_1^2}{\sig^2(1-\rho)}\times\frac{1+(K-u-1)\rho}{1+(K-1)\rho} \bs{1}_u \bs{1}'_u \rt| \\
%&=& 1+ \frac{2f(1-f)\beta_1^2}{\sig^2(1-\rho)}\frac{1+(K-u-1)\rho}{1+(K-1)\rho} u
%\end{eqnarray*}

So, $|\bI + \bH_u\bE^{-1}| -  |\bI + \bH_K\bE^{-1}| \overset{P} {\underset{n\rightarrow \infty}\rightarrow}
 \frac{2f(1-f)\beta_1^2}{\sig^2\{1+(K-1)\rho\}} \lt( \frac{1+(K-u-1)\rho}{1-\rho}u - K \rt) > 0$ under the condition
$\frac{u}{K} > \frac{\sig^2\{1-\rho\}}{\sig^2\{1+(K-u-1)\rho\}}$.
%This means, we expect the MANOVA statistic $|\bI + \bH\bE^{-1}|^{-1}$ under partial association to be closer to $0$ (i.e., more power) than that under complete association when the proportion of associated traits among the correlated traits exceeds the proportion of second largest eigenvalue of $\bSig_{22}$ to the largest one. 
%Here, $\bSig_{22}$ corresponds to the error covariance matrix $\bSig_{K-u}$ for the $K-u$ unassociated traits. 
It may be noted that the condition simplifies to $\rho>\frac{1}{u+1}$, which explains why we observe higher power for partial association and lower for complete association for $K=2$ traits once the within trait correlation $\rho$ exceeds $1/2$.
$\qed$

%%%%%%%
%\no\textbf
\subsubsection*{Proof of Theorem~\ref{M-Thm2} } 
Without loss of generality, let us assume that $\bY$ and $\bX$ are centered. 
%For testing $H_0: \bbeta=\bs{0}$, the Wilk's Lambda test statistic is $\text{det}\bE/\text{det}(\bH+\bE)=\text{det}(\frac{1}{n}\bE)/\text{det}(\frac{1}{n}\bH+\frac{1}{n}\bE)$ , where $\bH = \hat\bbeta (\bX'\bX) \hat\bbeta'$, $\bE = \bY'\bY - \hat\bbeta (\bX'\bX) \hat\bbeta'$, $n$ is the number of unrelated individuals, and $\hat\bbeta = \bY' \bX (\bX'\bX)^{-1}$ is the least squares estimate of the vector of genetic effects $\bbeta$. Note that $\bX'\bX = \sum_{i=1}^n X_i^2$ is a random variable (not a matrix), where $\E(X_i^2)=2f(1-f)=\Var(X_i) \: \forall \: i$.
% Using our distributional assumptions about centered $\bX$ and $\Eps$, it can be easily shown that 
%$ \frac{1}{n}\bH \overset{\P}\rightarrow 2f(1-f) \bbeta \bbeta'$ and
%$\frac{1}{n}\bE \overset{\P}\rightarrow \bSig$ as $n\rightarrow\infty$.
In particular, for $K=2$, 
$\frac{1}{n}\bH \overset{\P} \rightarrow
2f(1-f) \lt( \begin{matrix}  \beta_1^2  & \beta_1\beta_2 \\  \beta_1\beta_2 &  \beta_2^2 \end{matrix} \rt)$ and
$\frac{1}{n}\bE \overset{\P} \rightarrow
\sig^2 \lt( \begin{matrix}  1  & \rho \\  \rho &  1 \end{matrix} \rt)$ as $n\rightarrow\infty$.

Let us now consider the alternatives $H_{a1}: \beta_1\neq0, \beta_2=0$ (only 1 trait is associated), and $H_{a2}:\beta_1\neq\beta_2\neq0$ (both traits are associated).
Under $H_{a1}$, the $\bH/n$ matrix becomes 
$\frac{1}{n}\bH_1 \overset{P} \rightarrow 2f(1-f)  \lt( \begin{matrix} \beta_1^2  & 0  \\  0  &  0 \end{matrix} \rt) $ for large $n$. 
Let $\bH_2$ be the $\bH$ matrix under $H_{a2}$. So,
%\vspace{-8mm}
\begin{eqnarray*}
& &\text{det}\lt(\frac{\bH_1}{n}+\frac{\bE}{n}\rt) - \text{det}\lt(\frac{\bH_2}{n}+\frac{\bE}{n}\rt) \\
& \overset{P} \rightarrow& 
\lt| \begin{matrix} \sig^2 + 2f(1-f)\beta_1^2  & \rho\sig^2 \\ 
\rho\sig^2 & \sig^2 \end{matrix} \rt| - 
\lt| \begin{matrix} \sig^2 + 2f(1-f)\beta_1^2  & \rho\sig^2 + 2f(1-f)\beta_1\beta_2 \\ 
\rho\sig^2 + 2f(1-f)\beta_1\beta_2 & \sig^2 + 2f(1-f)\beta_2^2 \end{matrix} \rt|		\\
&=& 2f(1-f)\beta_2\sig^2(2\rho\beta_1-\beta_2)	\\
&>&0 \text{ if } \{ \beta_2<2\rho\beta_1 \text{ \& } \beta_2>0 \} \text{ or } \{ \beta_2>2\rho\beta_1 \text{ \& } \beta_2<0 \} 
\end{eqnarray*}
This means, we expect the statistic $|\bE|/|\bH_1+\bE|$ under $H_{a1}$  (when only 1 trait is associated) to be closer to $0$ than the statistic $|\bE|/|\bH_2 + \bE|$ under $H_{a2}$ when $\{ 0<\beta_2<2\rho\beta_1 \}$ or $\{0>\beta_2>2\rho\beta_1 \}$. Thus, for $K=2$, MANOVA is expected to have more power when 1 trait is associated than when both traits are associated if $0<\beta_2<2\rho\beta_1$ or $0>\beta_2>2\rho\beta_1$.
$\qed$

\newpage
\pagebreak
%%%%%%%%%  Web Appendix B
\subsection{}	\label{app2}

\paragraph{Acceptance Region for MANOVA based on $\bZ$}
\no Consider the MMLR model %in \eqref{mmlr} 
\begin{equation}
\bY_{n\times K} = \bX_{n\times1} \bbeta_{1\times K}' + \Eps_{n\times K} \label{mmlr}
\end{equation}
where $\bbeta'=(\beta_1,...,\beta_K)$ is the vector of fixed unknown genetic effects corresponding to the $K$ correlated traits, and 
$\Eps$ is the matrix of random errors. For testing that the SNP is not associated with any of the $K$ traits, the null hypothesis of interest is $H_0: \bbeta=\bs{0}$.

Assume $\Eps$ is a normal data matrix from $N_K(\bs{0},\bSig)$. The log-likelihood $l(\bbeta, \bSig)$ of the trait matrix $\bY$ is given by 
\begin{equation}
l(\bbeta, \bSig)  = -\frac{1}{2}n \log|2\pi\bSig| - \frac{1}{2} \text{tr} \lt\{ \bSig^{-1}  (\bY-\bX\bbeta')' (\bY-\bX\bbeta') \rt\} \label{loglik}
\end{equation}
where $\bSig$ is a positive definite matrix representing residual covariance among the traits. 
The MLE of $\bbeta$ and $\bSig$ are $\hat\bbeta=\bY' \bX (\bX'\bX)^{-1}$ and $\hat\bSig=\frac{1}{n} \bY'(\bI_K - \bX(\bX'\bX)^{-1} \bX')\bY$ respectively.
Under the null, $\bbeta=\bs{0}$ and the MLE of $\bSig$ is $\hat\bSig_0 = \frac{1}{n} \bY'\bY$. 
The likelihood ratio test (LRT) of $H_0$ based on the MMLR model with matrix normal errors is equivalent to MANOVA statistic $\bLam$ (Wilk's Lambda):
\begin{equation}
 - 2\log{\bLam} = 2 \lt(  l(\hat\bbeta, \hat\bSig) -  l(\bs{0}, \hat\bSig_0) \rt) 
= n \log{\frac{|\hat\bSig_0|}{|\hat\bSig|}} %= -n \log{\frac{|\bY'\bY|}{|\bY'\bY - \hat\bbeta (\bX'\bX)\hat\bbeta'|}}
= -n \log{\frac{|\bE|}{|\bH+\bE|}} \label{wilks}
\end{equation}
where $\bH$ and $\bE$ are the hypothesis and the error sum of squares and cross product (SSCP) matrices respectively.

Let us now consider the following notations: $\dot{\bl}(\bbeta) = \frac{\partial}{\partial \bbeta} l(\bbeta, \bSig)$; 
$\ddot{\bl}(\bbeta) = \frac{\partial^2}{\partial \bbeta^2} l(\bbeta, \bSig)$.  
The Fisher Information matrix under $H_0$ is $\bI(\bs{0}) = - \E_{\bbeta=\bs{0}}(\ddot{\bl}(\bbeta) )$.
Using Taylor's Expansion upto order 2, we can write the LRT statistic as
$$ - 2\log{\bLam}% = 2 \lt(  l(\hat\bbeta, \hat\bSig) -  l(\bs{0}, \hat\bSig_0) \rt) = 
= 2 \lt\{ 0 + \frac{1}{2} \sqrt{n} (\hat\bbeta - \bs{0})' \lt( - \frac{1}{n} \ddot{\bl}(\bbeta^*) \rt) (\hat\bbeta - \bs{0}) \rt\},
\text{ where } |\bbeta^* - \bs{0}| \leq |\hat\bbeta - \bs{0}|
$$
Observe that $\sqrt{n}(\hat\bbeta - \bs{0})' \overset{D}\rightarrow \bZ \sim N_K(\bs{0}, \bI^{-1}(\bs{0}))$. 
If a particular component of the true $\bbeta$ is large (small), we expect the corresponding component of $\hat\bbeta$ and hence of $\bZ$ to be large (small). Thus for $\bZ$ to be larger than $\bs{0}$, we need to have the true $\bbeta$ larger than $\bs{0}$.   
%We have the equivalence of MANOVA (LRT statistic) and the score test statistic :
We can then write the asymptotically equivalent form of MANOVA Wilk's Lambda statistic in terms of a statistic involving $\bZ$:
$$ - 2\log{\bLam} \overset{D}\rightarrow \bZ' \bI(\bs{0}) \bZ \overset{a}\sim \chi^2_K$$
Instead of drawing the acceptance region of Wilks Lambda statistic, one can draw the acceptance region of the test statistic $\bZ' \bI(\bs{0}) \bZ$. 
The ellipse representing acceptance region for MANOVA is asymptotically equivalent to
$$\Eps_c(\bz; \bS, \bar\bz)  \equiv \lt\{ \bz: (\bz-\bar\bz)' \bS^{-1} (\bz-\bar\bz) \leq c^2 \rt\} $$
where $\bS = (n-1)^{-1} \sum_{i=1}^n (\bz_i - \bar\bz)(\bz_i - \bar\bz)'$ and 
$c^2$ is the $95$-th percentile of the distribution of $\bZ$.
The boundary of the ellipse $\Eps_c$ is computed as a transformation of the unit circle, $\mathcal{U} = (\sin \theta, \cos \theta)$ for $\theta \in (0,2\pi)$.
Let $\bA = \bS^{1/2}$ be the Choleski square root of $\bS$ in the sense that $\bS = \bA\bA'$.
 Then,
$\Eps_c = \bar\bz + c\bA\mathcal{U}$ is an ellipse centered at the mean $\bar\bz=(\bar{z}_1,\bar{z}_2)$. The size of the ellipse reflects the standard deviations of $z_1$ and $z_2$ while the shape reflects their correlation.
$\bZ$ has a $N_K(\bs{0},\bI(\bs{0})^{-1})$ distribution due to which we expect $\bar\bz \approx \bs{0}$ and 
$\bS \approx \frac{1}{n} \sum \bz\bz' \overset{P}\rightarrow \bI(\bs{0})^{-1} = \frac{1}{2p(1+p)}\bSig$
where $p$ is the m.a.f. of the genetic variant. Thus, 
for drawing the theoretical acceptance region of MANOVA, we use the facts that
$\bar\bZ \overset{P}\rightarrow \bs{0}$ and $\bS \overset{P}\rightarrow \frac{1}{2p(1+p)} \bSig$.
For Figure 1 in the main manuscript, we assumed $\bSig = \sig^2 \{ (1-\rho)\bI_K + \rho \bs{1}\bs{1}'\}$ with $K=2$. 
The theoretical acceptance region for MANOVA will then be asymptotically equivalent to
$\Eps_c \lt(\bz; \frac{\bSig}{2p(1+p)}, \bs{0} \rt)  \equiv \lt\{ \bz: \bz' \lt( \frac{\bSig}{2p(1+p)} \rt)^{-1} \bz \leq c^2 \rt\} $.

\newpage
\pagebreak
%%%%%%%%%  Web Appendix C
\subsection{}	\label{app3}

%\emph{P-value calculation for USAT does not require independence}: \\
\paragraph{Details of the approximate p-value calculation for USAT}
\no Let $T_M %= -n \log{\frac{|\bE|}{|\bH+\bE|}}
= -2 \log \bLam
 \overset{a}\sim \chi^2_{K}$ be the MANOVA test statistic based on Wilk's lambda
and $T_S \overset{approx}\sim a\chi^2_d + b$ be the SSU test statistic based on score vector from marginal normal models.
For USAT, we first consider the weighted statistic $T_\omg = \omg T_M + (1-\omg)T_S$, where $\omg \in [0,1]$ is the weight. Both MANOVA and SSU are special cases of the class of statistics $T_\omg$. 
Under $H_0$, for a given weight $\omg$, $T_\omg$ is approximately a linear combination of chi-squared distributions. 
%For a given $\omg$, the p-value $p_\omg$ of the test statistic $T_\omg$ can be calculated using \cite{Liu09-chi} algorithm for chi-square approximation of non-negative quadratic forms. 
%It is worth noting that the calculation of $p_\omg$ does not require independence assumption of the two test statistics.
The computation of p-value $p_\omg$ of the test statistic $T_\omg$ does not require independence of the statistics $T_M$ and $T_S$.
A detailed explanation of the determination of $p_\omg$ is provided below.
%The p-value $p_\omg$ of the test statistic $T_\omg$ can be calculated using \cite{Liu09-chi} algorithm for chi-square approximation of non-negative quadratic forms. 

Observe that one can write $T_M = \bU' \bI(\bs{0})^{-1} \bU$, where $\bU$ is the score vector under $H_0: \bbeta=\bs{0}$ from the MMLR model \eqref{M-mmlr} and 
$\bI(\bs{0}) = -\E_{\bbeta=\bs{0}}\lt( \frac{\partial}{\partial \bbeta} l(\bbeta, \bSig) \rt) = \Cov(\bU) |_{\bbeta=\bs{0}}$ is the Fisher Information matrix under $H_0$. On the other hand, $T_S = \bU_M ' \bU_M$, where $\bU_M$ is the marginal score vector under $H_0$ from the marginal models in equation \eqref{M-marginal} of main paper. As derived in the main manuscript, $\bU_M = \bY'\bX/\hat\sig_0^2$, where $\bY$ is the $n\times K$ phenotype matrix, $\bX$ is the $n\times1$ genotype matrix and $\hat\sig_0^2$ is the MLE of $\sig^2$ %in model \eqref{M-marginal} 
under $H_0$. Similarly, one can show that $\bU = \hat\bSig_0^{-1} \bY'\bX$, where $\hat\bSig_0 = \bY'\bY/n$ is the MLE of $\bSig$ in MMLR model \eqref{M-mmlr} under $H_0$.
The estimated variance of the score vector $\bU$ under $H_0$ is given by $\Cov(\bU)|_{\bbeta=\bs{0}} = \bI(\bs{0}) = (\bX'\bX) \hat\bSig_0^{-1}$.
For a given weight $\omg$, one can thus write
\begin{eqnarray*}
T_\omg &=& \omg T_M + (1-\omg)T_S \\ %= \omg \bU' \bI(\bs{0})\bU + (1-\omg) \bU_M'\bU_M
&=& \omg \lt(\hat\bSig_0^{-1} \hat\sig_0^2 \bU_M\rt)' \bI(\bs{0})^{-1} \lt(\hat\bSig_0^{-1} \hat\sig_0^2 \bU_M \rt)
+ (1-\omg) \bU_M'\bU_M \\
&=& \bU_M' \lt( \omg \hat\sig_0^4 (\bX'\bX)^{-1} \hat\bSig_0^{-1} + (1-\omg)\bI_K \rt) \bU_M
\end{eqnarray*}
where $\bI_K$ is the identity matrix of order $K$.
Denote $\bA = \omg \hat\sig_0^4 (\bX'\bX)^{-1} \hat\bSig_0^{-1} + (1-\omg)\bI_K$, which is a $K\times K$ symmetric, non-negative definite matrix. 
Note that marginal score vector $\bU_M$ has mean $\bs{0}$, estimated variance $\Cov(\bU_M) = \bX'\bX \bY'\bY/(n\hat\sig_0^4)$, and has an asymptotic $K$-variate normal distribution.
Let $\bP$ be a $K\times K$ orthonormal matrix that converts 
$\bB = \Cov(\bU_M)^{1/2} \bA \Cov(\bU_M)^{1/2} = \omg \bI_K + (1-\omg) \Cov(\bU_M)$ 
to the diagonal form 
$\bGam = \text{diag}(\lam_1,...\lam_K)$, where $\lam_1 \geq 0, ..., \lam_K \geq 0$. %Let $r=rank(\bA) \leq K$.
The weighted statistic $T_{\omg}$ can, then, be expressed as a non-negative quadratic form:
 %linear combination of chi-squared distributions:
\begin{eqnarray}
T_\omg = \bU_M' \bA \bU_M = \bV_M' \bGam \bV_M = \sum_{j=1}^K \lam_j \chi^2_{h_j}(\del_j)	\label{Tw}
\end{eqnarray}
where $\bV_M = \bP \Cov(\bU_M)^{-1/2} \bU_M \overset{a}\sim N(\bs{0}, \bI_K)$, and $h_j = 1$, $\del_j=0$ for all $j=1,2,...,K$.
For a given $\omg \in [0,1]$, the p-value $p_\omg$ of the statistic $T_\omg$ can, thus, be calculated by %\cite{M-Liu09-chi}
Liu et al. (2009) algorithm as:
\begin{eqnarray}
p_\omg = 1 - \pr\lt(T_\omg > t_\omg \rt) \approx 1 - \pr\lt( \chi^2_l(\del) > t_\omg^* \sig_\chi + \mu_\chi  \rt)	\label{pw}
\end{eqnarray}
where $t_\omg$ is the observed value of $T_\omg$ statistic, $t_\omg^* = (t_\omg - \E(T_\omg))/\sqrt{\Var(T_\omg)}$,
$\mu_\chi = \E\lt(\chi^2_l(\del) \rt) = l + \del$, $\sig_\chi = \sqrt{\Var\lt(\chi^2_l(\del) \rt)} = \sqrt{2(l+2\del)}$. The parameters $\del$ and $l$ are chosen such that the skewness of $T_\omg$ and $\chi^2_l(\del)$ are same and the difference between the kurtoses of $T_\omg$ and $\chi^2_l(\del)$ is minimized.

%The power of the test based on $T_\omg$ is maximized when $\omg$ is chosen appropriately. 
Apriori the optimal weight $\omg$ is not known. % and needs to be estimated from the data.
% by selecting the value of $\omg$ that maximizes power. 
%In other words, $T_{opt}=\{T_\omg : \min_{0\leq\omg\leq1} p_\omg \}$.
We propose our unified test USAT as $$T_{USAT}=\min_{0\leq\omg\leq1} p_{\omg} $$
%In other words, the optimal p-value is $\min_{0\leq\omg\leq1} p_\omg$. 
Thus, the USAT test statistic is not exactly the best weighted combination of MANOVA and SSU. It is the minimum of the p-values of the different weighted combinations. 
%Lee et al. (2012) proposed a similar test statistic based on minimum p-value in the context of rare variants in sequencing association studies.
For practical implementations of USAT, a grid of $11$ $\omg$ values were considered: $\{\omg_1=0,\omg_2=0.1,...,\omg_{10}=0.9,\omg_{11}=1\}$. 
%A finer grid of more $\omg$ values did not change the USAT power curve much. 

To find the p-value of our USAT test statistic, we need the null distribution of USAT.
%One option is to calculate the empirical p-value by considering several permuted datasets or by generating several datasets under the null (as done for Figure \ref{M-fig4} in main paper). Finding empirical p-values is computationally intensive and is not suitable when USAT is applied on a GWAS scale with large number of traits. 
We propose an approximate p-value calculation using a one-dimensional numerical integration, which makes USAT suitable for application on a GWAS scale. Observe that the p-value of statistic $T_{USAT}$ is
\allowdisplaybreaks
\begin{eqnarray*}
p_{USAT}&=&1-P(T_{USAT}\leq t_{USAT})\\
&=& 1-P\lt(\min_\omg{p_\omg} \leq t_{USAT}\rt) = 1-P\lt(1-\min_\omg{p_\omg} > 1-t_{USAT}\rt) \\
&=& 1-P\lt(\max_\omg{(1-p_\omg)} > 1-t_{USAT}\rt) \\
&=& 1-P\lt( \{1-p_{\omg_1} > 1-t_{USAT}\}, \ldots, \{1-p_{\omg_{11}} > 1-t_{USAT}\}  \rt) \\
&=& 1- P\Big( \{(1-p_{\omg_1})^{th} \text{ quantile} < (1-t_{USAT})^{th} \text{ quantile} \}, \ldots, \\
& & \{(1-p_{\omg_{11}})^{th} \text{ quantile} < (1-t_{USAT})^{th} \text{ quantile} \}  \Big) \\
&=&1-P\lt(T_{\omg_1}<q_{\min}(\omg_1),..., T_{\omg_{11}}<q_{\min}(\omg_{11})\rt) \\
&=& 1 - \int F_{T_S|T_M}\big(\del_\omg(x)|x \big) f_{T_M}(x) dx \\
&\approx& 1 - \int_{0}^{\infty} F_{T_S}\big(\del_\omg(x)|x \big) f_{T_M}(x) dx
\end{eqnarray*}
where $t_{USAT}$ is the observed value of USAT test statistic for a given dataset, \\
 $q_{\min}(\omg_b)$ is the $(1-t_{USAT})$-th percentile of the distribution of $T_{\omg_b}$ for a given $\omg=\omg_b$, \\
$F_{T_S|T_M}(.|T_M=x)$ is the conditional cdf of SSU statistic $T_S$ given MANOVA statistic $T_M$, \\
$F_{T_S}(.)$ is the cdf of SSU test statistic $T_S$, \\
$\del_\omg(x)=\min_{\omg \in \{\omg_1,...,\omg_{11}\}} \frac{q_{\min}(\omg)-\omg x}{1-\omg}$, \\
and  $f_{T_M}(.)$ is the pdf of MANOVA test statistic $T_M$. \\

%\no\emph{Computation of the integral: } 
For the integral $\int_0^{\infty} F_{T_S}\big(\del_\omg(x) \big) f_{T_M}(x) dx$, we first need to evaluate %$F_{T_S}\big(\del_\omg(x) \big) = P(T_S \leq \del_\omg(x))$.
$$
F_{T_S}\big(\del_\omg(x) \big) = \pr \lt( T_S \leq  \del_\omg(x) \rt) \approx \pr \lt( a\chi_d^2 + b \leq \del_\omg(x) \rt)
= \pr \lt( \chi_d^2 \leq \frac{\del_\omg(x) - b}{a} \rt)
$$
This can be easily evaluated using \texttt{R} function \texttt{pchisq()}.
% Observe that the distribution of the SSU test statistic $T_S = \bU_M' \bU_M$ is a weighted sum of independent chi-squared variables with $df=1$ (as evident from equation \eqref{Tw} with $\omg=0$). Denote this weighted combination as $Q$ where $c_j$ is the $j$-th ordered eigenvalue of $\Cov(\bU_M)$:
%$$ Q =  \sum_{j=1}^K c_j \chi_1^2$$
%Thus, $ \displaystyle F_{T_S}\big(\del_\omg(x) \big) = \pr \lt( Q \leq \del_\omg(x) \rt) = 
%\pr \lt( \frac{Q-\mu_Q}{\sig_Q} \leq \frac{\del_\omg(x) - \mu_Q}{\sig_Q} \rt)$
The integrand as a function of $x$ can then be coded as
\texttt{pchisq((delta.x-b)/a, df=d, ncp=0)*dchisq(x, df=K)}.
The integration has been performed numerically using \texttt{R} function \texttt{integrate()}. When the optimal choice of $\omg$ lies near the boundary (i.e., close to $0$ or $1$) and the corresponding statistic ($T_S$ or $T_M$ depending upon whether optimal $\omg$ is close to $0$ or $1$) is highly significant (i.e., corresponding p-value is of the order of $10^{-8}$), the function \texttt{integrate} can have low accuracy and can give rise to an integral value exceeding 1. In such a scenario, \texttt{R} function \texttt{quadinf()} from package \texttt{pracma} (Borchers, 2012) can give very accurate results. The cost of accuracy is longer computation time: \texttt{quadinf} takes almost twice as much time compared to \texttt{integrate}. For our simulated datasets as well as real dataset, we found the two functions 
%\texttt{integrate()} and \texttt{quadinf()} 
giving very similar results in most situations except in the afore-mentioned scenario where \texttt{integrate} gave negative p-values for USAT. In such rare situations, we implemented the numerical integration using \texttt{quadinf}.

\subsubsection{References:}
Borchers, H.W. (2012). pracma: Practical Numerical Math Functions. R
  package version 0.9.6. http://CRAN.R-project.org/package=pracma. \\
%Lee, S., Wu, M., and Lin, X. (2012). Optimal tests for rare variant effects in sequencing association studies. \emph{Biostatistics} \textbf{13,} 762-775. \\
Liu, H., Tang, Y., and Zhang, H. (2009). A new chi-square approximation to the distribution of non-negative definite quadratic forms in non-central normal variables. \emph{Comput Stat Data Anal} \textbf{53,} 853-856. \\
R Development Core Team (2011). R: A language and environment for
  statistical computing. R Foundation for Statistical Computing, Vienna, Austria. ISBN 3-900051-07-0, URL http://www.R-project.org/.

\newpage
\pagebreak
%%%%%%%%%  Web Appendix D
\subsection{}	\label{app4}

\paragraph{Simulation 4: Other correlation structures}
\no Apart from the compound symmetry (CS) structure, we also considered AR1($\rho$) and other structures for correlation in our simulation studies. Details on how the datasets were simulated can be found in Section 3 of our main paper.

\begin{figure}[h!]
	\begin{center}
	\includegraphics[width=6in,height=3in]{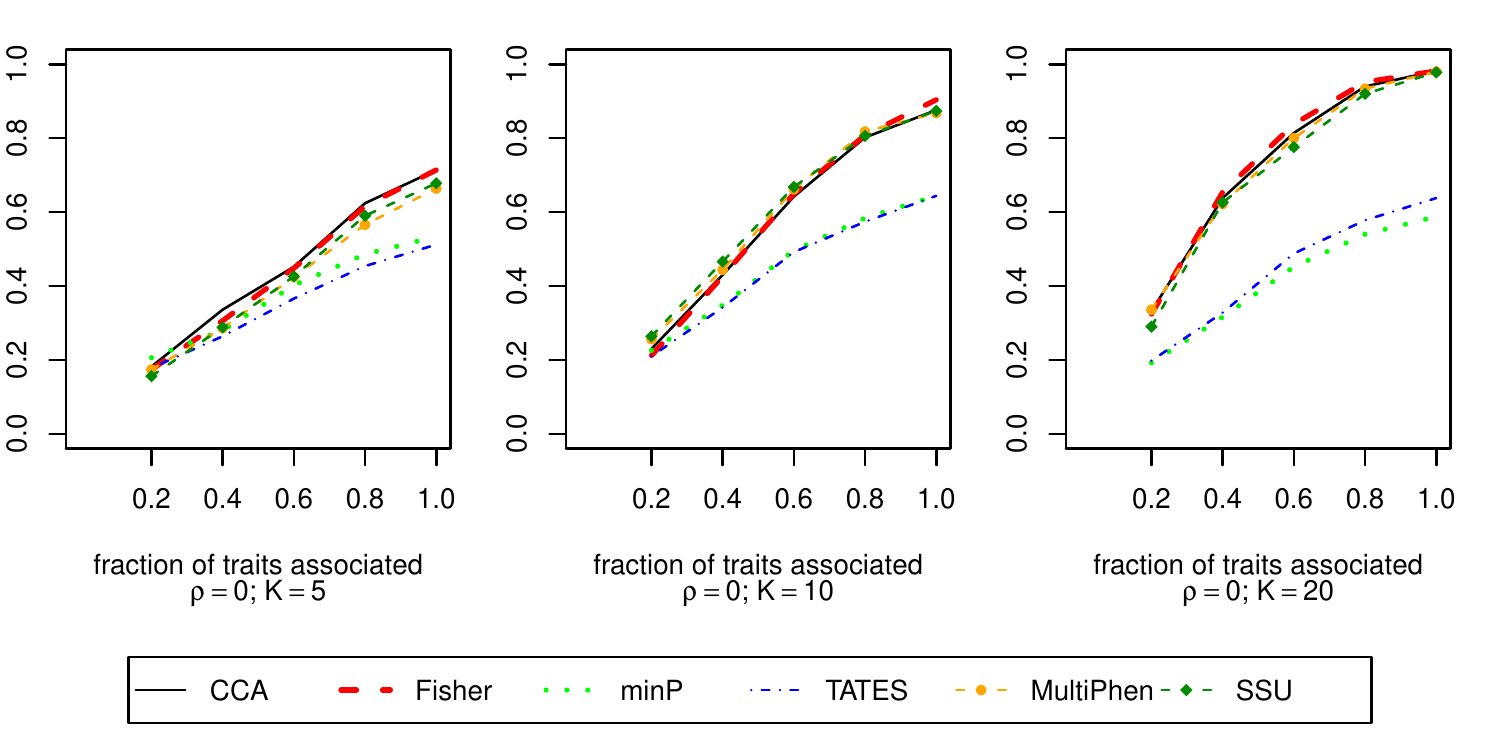}
	\caption[Simulation 4 power curves: 5,10,20 traits, correlation structure I (uncorrelated)]{\textbf{Correlation structure I (uncorrelated)}: Empirical power curves of the different association tests for $K=5,10,20$ traits and within trait correlation $\rho=0$ based on $N=500$ datasets. The correlation structure assumes all traits to be uncorrelated. Same direction and same size effects (effect size of $0.395$) are used when 2 or more traits are associated. The power is plotted along y-axis while the fraction of traits associated with the genetic variant is plotted along x-axis.}
	\label{fig4i}
	\end{center}
\end{figure}

\emph{Correlation Structure I: uncorrelated traits }: We assumed that none of the traits was correlated with another. From Figure \ref{fig4i}, we see that performances of all methods are similar except minP/TATES. 
All the methods, including MANOVA, have steadily rising power curves with increase in proportion of associated traits. This confirms that  MANOVA's lack of power in detecting pleiotropy in certain situations is primarily due to the correlatedness of all the traits.

\begin{figure}[t!]
	\begin{center}
	\includegraphics[height=7.5in]{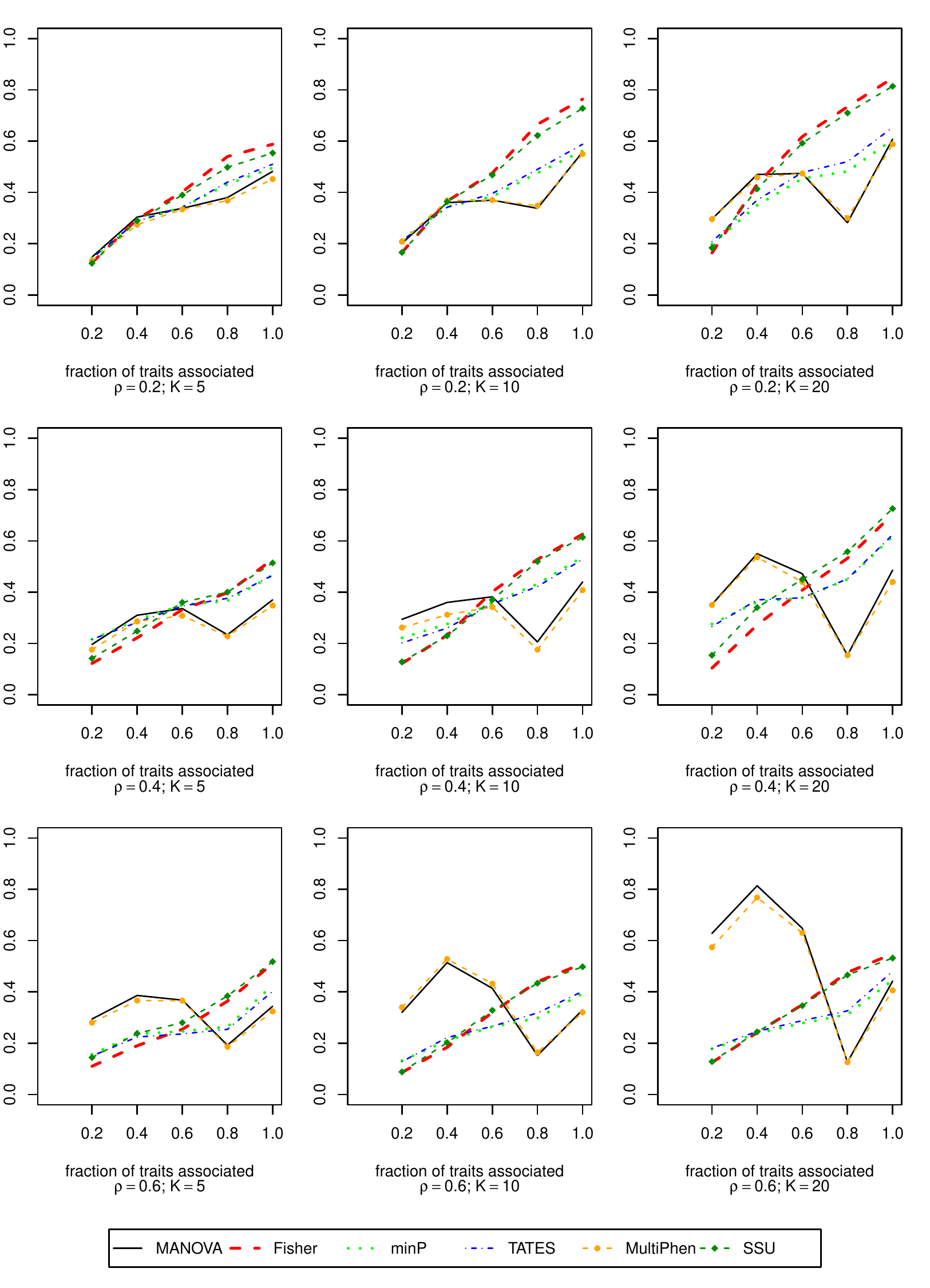}
	\caption[Simulation 4 power curves: 5,10,20 traits, correlation structure II]{\textbf{Correlation structure II}: Empirical power curves of the different association tests for $K=5,10,20$ traits and different within trait correlation values $\rho=0.2,0.4,0.6$ based on $N=500$ datasets. This correlation structure assumes that the first $80\%$ of the traits are correlated (Compound Symmetry structure with correlation $\rho$) and the last $20\%$ of the traits are independent of the others. Same direction and same size effects (effect size of $0.395$) are used when 2 or more traits are associated. The power is plotted along y-axis while the fraction of traits associated with the genetic variant is plotted along x-axis. Upto the point $0.8$ on the x-axis, all the traits are correlated.}
	\label{fig4ii}
	\end{center}
\end{figure}

\emph{Correlation Structure II}: Here we assumed that first $80\%$ of the $K$ traits were correlated (with a compound symmetry structure) and the rest $20\%$ were uncorrelated. 
For our simulation study, we considered $K=5,10,20$ traits and positive correlation parameter $\rho=0.2,0.4,0.6$. 
%We did not consider $\rho=0.8$ since such high pairwise correlation is not expected when $K$ is not small.
In such a situation we noticed that as correlation increased among the associated traits, the power of MANOVA dropped. 
Figure \ref{fig4ii} shows that the lowest point in the MANOVA power curve occurs at $0.8$ on the axis, which means MANOVA has the least power in detecting association when all the correlated traits are associated. At point $1.0$ on the x-axis, when all the traits are associated but not all are correlated, the performance of MANOVA improves but not as good as the methods that do not explicitly consider the covariance matrix in the test statistic. %In the latter halves of the plots, performance of the other methods (Fisher's, minP/TATES, SSU) are better than MANOVA or MultiPhen. 

An important observation from Figure \ref{fig4ii} is that MANOVA is not expected to suffer from power loss at `complete association' (when all traits are associated) if all associated traits are not correlated (refer \ref{app5} %Web Appendix S5 
for theoretical result).

\begin{figure}[h!]
	\begin{center}
	\includegraphics[width=6in,height=7.8in]{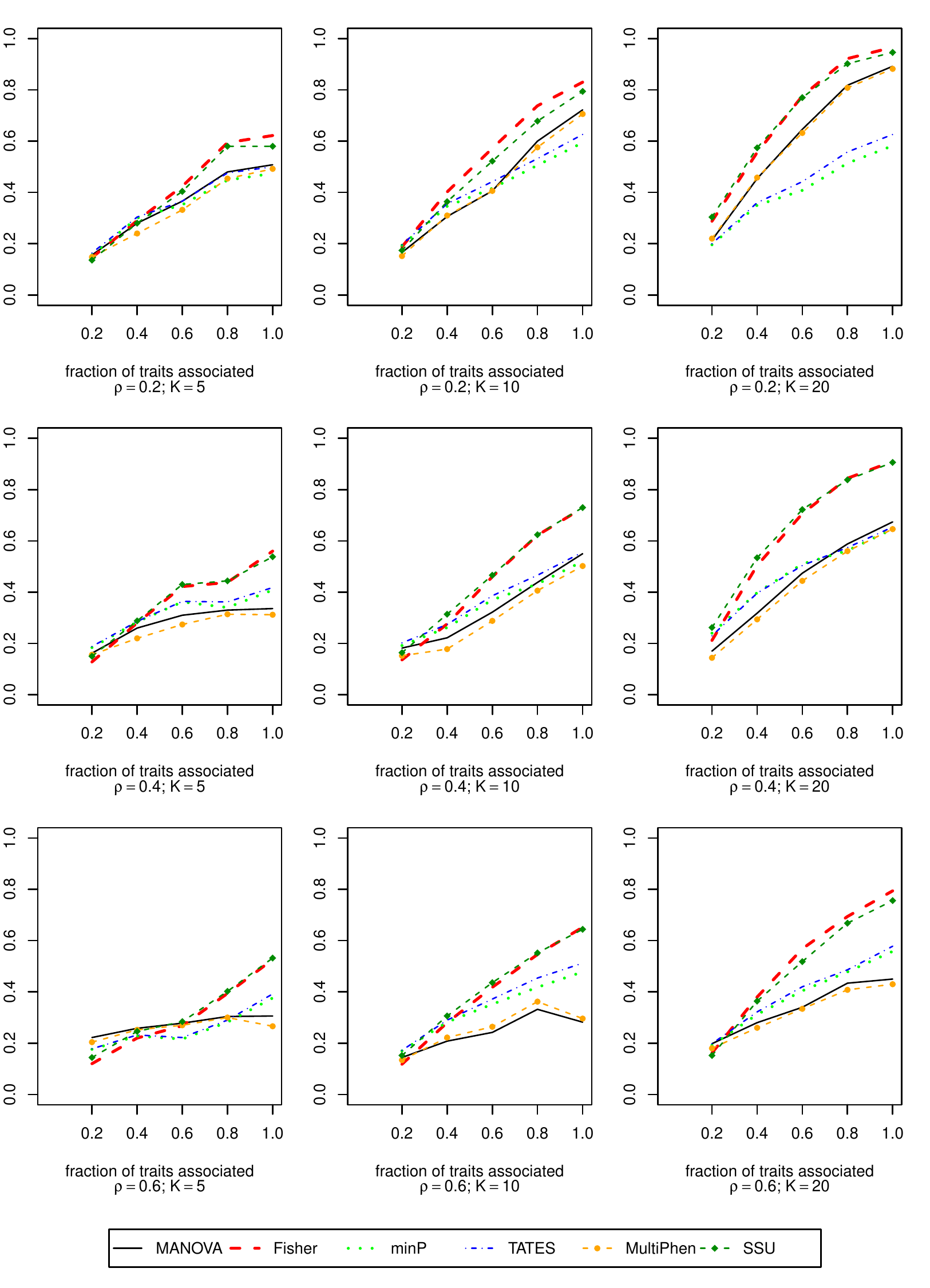}
	\caption[Simulation 4 power curves: 5,10,20 traits, correlation structure III (AR1)]{\textbf{Correlation structure III (AR1)}: Empirical power curves of the different association tests based on $N=500$ datasets for $K=5,10,20$ traits and AR1($\rho$) correlation structure with $\rho=0.2,0.4,0.6$. Same direction and same size effects (effect size of $0.395$) are used when 2 or more traits are associated. The power is plotted along y-axis while the fraction of traits associated with the genetic variant is plotted along x-axis.}
	\label{fig4iii}
	\end{center}
\end{figure}

\vspace{0.8cm}
\emph{Correlation Structure III: AR1($\rho$) }: For given $K$ traits, we assumed the covariance structure
 $\bSig = \sig^2 \bR(\rho) = \sig^2 \lt( 
\begin{array}{ccccc} 
1 & \rho & \rho^2 & \ldots & \rho^{K-1} \\ 
\rho & 1 & \rho & \ldots & \rho^{K-2} \\
\vdots & &  & \ddots & \vdots \\
\rho^{K-1} & \rho^{K-2} & \rho^{K-3} & \ldots & 1
\end{array} \rt) $. Figure \ref{fig4iii} shows that for a given $\rho$, MANOVA performs better with increase in $K$ and with increase in the fraction of associated traits. This is so because at a higher fraction (on the x-axis), the AR1 correlation among traits becomes negligible and the latter traits are effectively uncorrelated (the behavior we saw in Figures \ref{fig4i} \& \ref{fig4ii}). 
Observe that for a given $\rho$, the power at or near `complete association' (where all traits are associated) increases with increase in $K$ since for the latter traits, the correlation rapidly goes towards $0$.
With increase in the parameter $\rho$ and for small $K$, we start observing  MANOVA's lack of power as the latter pairwise correlations are not effectively zero.

\newpage
\pagebreak
%%%%%%%%%  Web Appendix E
\subsection{}	\label{app5}

%\paragraph{} 
Figure \ref{fig4ii} shows that if all the traits are not correlated, MANOVA does not experience power loss for testing $H_0$ even when all the traits are associated. This behavior is theoretically explained by the following theorem for the special case of CS residual correlation structure for the correlated traits.

\begin{theorem*}
Without loss of generality, let $\bY$ and $\bX$ be the centered phenotype matrix and the centered genotype vector respectively.
Consider the MMLR model 
$$\bY_{n\times K} = \bX_{n\times1} \bbeta_{1\times K}' + \Eps_{n\times K}, \:\: vec(\Eps)\sim N_{nK}(\bs{0}, \bI_n\otimes\bSig)$$ 
where 
$\bSig_{K\times K} = \lt( \begin{matrix} \bSig_{11 (m\times m)} & \bSig_{12} \\ \bSig_{21} & \bSig_{22}
\end{matrix}\rt)$, 
$\bSig_{11}=\sig^2 \lt( (1-\rho)\bI_m + \rho \bs{1}\bs{1}' \rt)$, 
$\sig^2>0$, $\rho \:(>0)$ is the within trait correlation such that $\bSig_{11}$ is a positive definite covariance matrix,
$\bSig_{12}=\bSig_{21}'=\bO_{m\times (K-m)}$,
$\bSig_{22}=\sig^2 \bI_{K-m}$
and $\bbeta'=(\beta_1,...,\beta_K)$ is the vector of genetic effects.
Assume that the genetic effects of the associated traits are equal in size and positive. 
Consider two scenarios of association: `partial association' (when the SNP is associated with $u$ $(<K)$ traits), and `complete association' (when all $K$ traits are associated).

For testing $H_0: \bbeta=\bs{0}$, MANOVA is not expected to suffer from power loss at `complete association' compared to `partial association' with $u \:(>m)$ associated traits.
\end{theorem*}

%\underline{\textbf{Proof}}: 
\begin{proof}
%\begin{proof} \let\qed\relax
For the $m\times m$ CS residual covariance sub-matrix $\bSig_{11}$, we know that the eigen vector corresponding to the largest eigenvalue $\lam_{(m)1}=\sig^2\{1+(m-1)\rho\}$ is $\bv_1 \propto \bs{1}$, while the eigen vectors corresponding to $\lam_{(m)2}=...=\lam_{(m)m}=\sig^2(1-\rho)$ are respectively $\bv_2,...,\bv_m$ such that $\bs{1}'\bv_k=0 \: \forall \: k=2,...,m$. For the eigen vectors to be orthonormal, we must have $\bv_1=c_m \bs{1}$ such that $\sqrt{c_m^2+...+c_m^2}=1 \iff c^2_m=1/m$.
Thus, we can write,
$$\bSig_{11 (m\times m)} =\lam_{(m)1} c^2_m \bs{1}\bs{1}' + \sum_{i=2}^m \lam_{(m)i} \bv_i \bv'_i \text{ and } 
\bSig_{11}^{-1} = \frac{1}{\lam_{(m)1}} c^2_m \bs{1}\bs{1}' + \sum_{i=2}^m \frac{1}{\lam_{(m)i}} \bv_i \bv'_i$$

Consider the 2 alternatives $H_{a,u}: \beta_1=...=\beta_u\neq0, \beta_{K-u}=...=\beta_K=0$ (partial association) and $H_{a,K}: \beta_1=...=\beta_K\neq0$ (complete association) against the null hypothesis $H_0: \beta_1=...=\beta_K=0$. Here, for the partial association case, $u \:(>m)$ is the number of traits associated and $m$ is the number of correlated traits. 
%So, $K-m$ traits out of $K-u$ unassociated traits are uncorrelated.
In the following, the notation $\overset{P}\rightarrow$ denotes convergence in probability as $n\rightarrow\infty$.

Under the alternative $H_{a,K}$ (complete association), it can be shown that
\begin{eqnarray*}
%|\bI + \bH\bE^{-1}| \overset{H_{a,K}}{:=} 
\lt|\bI + \frac{\bH_K}{n}\lt(\frac{\bE}{n}\rt)^{-1}\rt| 
	&\overset{P}\rightarrow&
\lt| \bI_K + (2pq \beta_1^2) 
\lt( \begin{matrix} \bs{1}_m\bs{1}_m' &  \bs{1}_m\bs{1}_{K-m}' \\
\bs{1}_{K-m} \bs{1}_m' &  \bs{1}_{K-m}\bs{1}_{K-m}'
\end{matrix} \rt)
\lt( \begin{matrix} \bSig_{11}^{-1} & \bO \\ \bO & \frac{1}{\sig^2}\bI_{K-m}
\end{matrix}\rt) \rt| \\
%	&=& \lt| \bI_K + (2pq \beta_1^2) 
%\lt( \begin{matrix} \bs{1}_m\bs{1}_m' &  \bs{1}_m\bs{1}_{K-m}' \\
%\bs{1}_{K-m} \bs{1}_m' &  \bs{1}_{K-m}\bs{1}_{K-m}'
%\end{matrix} \rt)
%\lt( \begin{matrix}
%\frac{1}{\lam_{(m)1}} c^2_m \bs{1}_m\bs{1}_m' + \sum_{i=2}^m \frac{1}{\lam_{(m)i}} \bv_i \bv'_i & \bO \\
%\bO & \frac{1}{\sig^2}\bI_{K-m}
%\end{matrix} \rt)  \rt| \\
%	&=& \lt| \lt( \begin{matrix} \bI_m & \bO \\ \bO & \bI_{K-m} \end{matrix} \rt)
%+ (2pq \beta_1^2) 
%\lt( \begin{matrix}
%\frac{1}{\lam_{(m)1}} \bs{1}_m\bs{1}_m' & \frac{1}{\sig^2} \bs{1}_m\bs{1}_{K-m}' \\
%\frac{1}{\lam_{(m)1}} \bs{1}_{K-m}\bs{1}_m' & \frac{1}{\sig^2} \bs{1}_{K-m}\bs{1}_{K-m}' 
%\end{matrix} \rt)  \rt| \\
%	&=& \lt| \begin{matrix} 
%\bI_m + a \bs{1}_m\bs{1}_m' & b \bs{1}_m\bs{1}_{K-m}' \\ 
%a \bs{1}_{K-m}\bs{1}_m' & \bI_{K-m} + b \bs{1}_{K-m}\bs{1}_{K-m}'
%\end{matrix} \rt|, 
%\text{ where } a=\frac{2pq\beta_1^2}{\sig^2\{1+(m-1)\rho \}},
%b=\frac{2pq\beta_1^2}{\sig^2} \\
	&=& \lt| (\bI_m + a \bs{1}_m\bs{1}_m') - 
( b \bs{1}_m\bs{1}_{K-m}')(\bI_{K-m} + b \bs{1}_{K-m}\bs{1}_{K-m}')^{-1}(a \bs{1}_{K-m}\bs{1}_m' ) \rt| \\
&& \times |\bI_{K-m}+b\bs{1}\bs{1}'|  \\
%	&=& \lt| (\bI_m + a \bs{1}_m\bs{1}_m') - 
%\lt( b \bs{1}_m\bs{1}_{K-m}' - bc (K-m) \bs{1}_m \bs{1}_{K-m}' \rt) (a \bs{1}_{K-m}\bs{1}_m' ) \rt| \times |\bI_{K-m}+b\bs{1}\bs{1}'| \\
%	&=&  \lt| (\bI_m + a \bs{1}_m\bs{1}_m') - 
%( c \bs{1}_m\bs{1}_{K-m}' ) (a \bs{1}_{K-m}\bs{1}_m' ) \rt| \times |\bI_{K-m}+b\bs{1}\bs{1}'|,
%\text{ since } bc(K-m)=b-c \\
	&=&  \lt| (\bI_m + a \bs{1}_m\bs{1}_m') - ac(K-m) \bs{1}_m\bs{1}_m' \rt| 
\times |\bI_{K-m}+b\bs{1}\bs{1}'| \\
%	&=& \lt( 1 + [a-ac(K-m)]m \rt) \times [1+b(K-m)] \\
%	&=& \{1+am-acm(K-m)\} \times \{1+b(K-m)\} \\
%	&=& 1 + b(K-m) + am + abm(K-m) - acm(K-m) - abcm(K-m)^2 \\
	&=& 1 + b(K-m) + am%, \text{ since } am(K-m) \{ c + bc(K-m) \} = abm(K-m)
\end{eqnarray*}
where  
$a=\frac{2pq\beta_1^2}{\sig^2\{1+(m-1)\rho \}}$,
$ b=\frac{2pq\beta_1^2}{\sig^2}$, 
$ (\bI_{K-m} + b \bs{1}_{K-m}\bs{1}_{K-m}')^{-1} = \bI - c \bs{1}\bs{1}'$, 
$c = \frac{b}{1+(K-m)b}$.

For $u(>m)$ associated traits, let us now partition the residual covariance matrix as
$$\bSig_{K \times K} = \lt( \begin{matrix} \bS_{11 (u\times u)} & \bS_{12} \\
\bS_{12}' & \bS_{22} \end{matrix} \rt)
\text{ where } \bS_{11} = \lt( \begin{matrix} \bSig_{11 (m\times m)} & \bO \\
\bO & \sig^2 \bI_{u-m} \end{matrix} \rt),
\:\: \bS_{12} = \bO_{u\times(K-u)}, \:\: \bS_{22}=\sig^2 \bI_{K-u}
$$

Under the alternative $H_{a,u}$ (partial association) where $0<m<u<K$, one can show that
\begin{eqnarray*}
\lt|\bI + \frac{\bH_u}{n}\lt(\frac{\bE}{n}\rt)^{-1}\rt| 
	&\overset{P}\rightarrow&
\lt| \bI_K + 2pq 
\lt( \begin{matrix} \beta_1^2 \bs{1}_u\bs{1}_u' &  \bO \\
\bO &  \bO  \end{matrix} \rt)
\lt( \begin{matrix} \bS_{11}^{-1} & \bO \\ \bO & \frac{1}{\sig^2}\bI_{K-u}
\end{matrix}\rt) \rt| \\
%	&=& \lt| \bI_K + 2pq 
%\lt( \begin{matrix} \beta_1^2 \bs{1}_u\bs{1}_u' \bS_{11}^{-1} &  \bO \\
%\bO &  \bO  \end{matrix} \rt) \rt| \\
%	&=& \lt| \begin{matrix} \bI_u + 2pq \beta_1^2 \bs{1}_u\bs{1}_u' \bS_{11}^{-1} &  \bO \\
%\bO &  \bI_{K-u}  \end{matrix} \rt| \\
	&=& \lt|  \bI_u + 2pq \beta_1^2 \bs{1}_u\bs{1}_u'  
\lt( \begin{matrix} \bSig_{11}^{-1} &  \bO \\
\bO &  \frac{1}{\sig^2}\bI_{u-m}  \end{matrix} \rt) \rt| \\
	&=& 1+b(u-m)+am, \text{ where } a=\frac{2pq\beta_1^2}{\sig^2\{1+(m-1)\rho \}},
b=\frac{2pq\beta_1^2}{\sig^2}
\end{eqnarray*}
$$
 \therefore \:\: \lt|\bI_K + \bH_K \bE^{-1}\rt| -  \lt|\bI_K + \bH_u \bE^{-1}\rt| 
	\overset{P}\rightarrow  b(K-u) > 0 
$$
\end{proof}

\newpage
\pagebreak
%%%%%%%%%  Web Appendix F
\subsection{}	\label{app6}

\paragraph{Details on ARIC Study phenotypes and covariate choices}
\no ARIC has collected %fasting glucose measures along with other 
measures on many type 2 diabetes (T2D) related traits at 4 separate visits over a 9-year period. 
A diagnosis of T2D is considered positive if fasting plasma glucose concentration is $\geq 126$ mg/dL, or casual plasma glucose level is $\geq 200$ mg/dL, or 2-hour plasma glucose value after a standard glucose challenge is $\geq 200$ mg/dL %\citep{WHO03}. 
(WHO, 2003). All analytes were determined at central laboratories according to standard protocols: plasma glucose by a hexokinase assay, and insulin by radioimmunoassay ($^{125}$Insulin Kit; Cambridge Medical Diagnosis, Billerica, MA).
Sedentary lifestyle and obesity are major risk factors for T2D. In addition to general obesity, the distribution of body fat (or abdominal obesity, as estimated by waist-to-hip circumference ratio) contributes to T2D risk.
For our analysis, we focused on the Caucasian participants and the following 3 T2D related quantitative traits %or risk factors 
measured at visit $4$ ($1996-98$): fasting glucose;  2-hour glucose from an oral glucose tolerance test; fasting insulin. %; BMI; waist circumference.  
The pairwise correlations among these 3 traits were within $(0.2,0.35)$. 
These traits are substantially affected by treatment with diabetes medications, and so statistical analysis results are not generally interpretable in the same way they can be interpreted in non-diabetic individuals.
Other available traits were Body Mass Index (BMI) and waist circumference. BMI was calculated as weight/height$^2$ (kg/m$^2$), and obesity was defined as a BMI $\geq 30$ kg/m$^2$. Waist circumference was measured at the umbilical level. 
Due to a high pairwise correlation of $0.9$ between waist circumference and BMI,  we chose only BMI. However, as in most studies of T2D, BMI was used as a covariate along with age and sex.
%Figure \ref{fig5a} shows the pairwise correlations among these 5 traits, all of which lie within $(0.2,0.9)$. 
%The highest correlation is observed between the traits BMI and waist circumference. 
Individuals with diagnosed or treated diabetes
%(report of physician diagnosis or use of hypoglycemic medications)
 at visit $4$ and individuals with missing traits were excluded, leaving $5,816$ in our analytic sample. 
%Statistical models were adjusted for Age, Sex and BMI.

\subsubsection{Reference:}
WHO (2003). Screening for Type 2 Diabetes: Report of a World Health Organization and
International Diabetes Federation meeting. \emph{Geneva}.

\newpage
\pagebreak
%%%%%%%%%  Web Appendix G
\subsection{}	\label{app7}

\paragraph{Covariate Adjustment for USAT}
\no The ARIC data analysis using USAT required covariate adjustment (predictors other than SNP). This version of USAT requires covariate adjustment for both SSU test and MANOVA. Once the adjusted MANOVA and SSU test statistics are available, one can easily compute approximate p-value for USAT (refer section \ref{M-sec:usat} of the main paper for the p-value calculation method). Let $\bZ_{n\times q}$ be the matrix of $q$ covariates (other than SNP) for $n$ unrelated individuals. Without loss of generality, the phenotype matrix $\bY$, the genotype vector $\bX$ and the covariate matrix $\bZ$ are centered (but not scaled). The following paragraphs outline the details of such covariate adjustment.\\

\paragraph{MANOVA with covariate adjustment} 
The MMLR model for the association test of $K$ traits and the SNP (after adjusting for other covariates):
$$
\bY_{n\times K} = \bX_{n\times1} \bbeta_{1\times K}' + \bs{1}' \otimes \bZ\bPhi + \Eps_{n\times K}
$$
where $\bbeta'=(\beta_1,...,\beta_K)$ is the vector of fixed unknown genetic effects corresponding to the $K$ correlated traits, and 
$\Eps$ is the matrix of random errors. 
For testing that the SNP is not associated with any of the $K$ traits, the null hypothesis of interest is $H_0: \bbeta=\bs{0}$.
For testing $H_0$, the LRT is equivalent to the MANOVA test statistic, which is the ratio of generalized variances 
$\bLam = {|\bE|}/{|\bH+\bE|}$. 
%Here, $\bH+\bE$ is calculated as the covariance matrix of the residuals obtained after fitting the model under $H_0$. $\bE$ is calculated as the covariance matrix of the residuals of the model after adjusting for the $q$ covariates. In other words,
Here, $\bH+\bE$ is the covariance matrix of the $K$ residual vectors where the $k$-th residual vector is obtained by fitting the model for $k$-th trait under $H_0$. $\bE$ is the covariance matrix of the $K$ residual vectors where the $k$-th residual vector is obtained by fitting the full model for $k$-th trait. Under $H_0$,
 Wilk's Lambda $-2\log\bLam$ has an approximate asymptotic $\chi^2_K$ distribution under $H_0$. \\

%%% Our new version of SSU %%%%
%\subsubsection*{SSU test based on $\bU$ from marginal normal model with covariate adjustment} \label{ssu-marg-n}
\no \emph{SSU Test with covariate adjustment}

\no For $k$-th trait vector, we assume the marginal normal model :
$$ \bY_k = \beta_k \bX + \bZ \bPhi + \beps_k, \:\: \beps_k \sim N_n(\bs{0}, \sig^2\bI_n) $$
$\beta_k$ is the parameter associated with the SNP effect on the $k$-th trait. $\bPhi$ is the $q\times 1$ vector of parameters associated with the $q$ covariates. The null hypothesis associated with $k$-th marginal model is $H_{0k}: \beta_k=0$. We need to obtain the MLE $\hat\bPhi$ under the global null $H_0: \cap_{k=1}^K H_{0,k}$.
Under $H_{0,k}$, the $k$-th marginal model is 
$$\bY_k = \bZ \bPhi + \beps_k, \:\: \beps_k \sim N_n(\bs{0}, \sig^2\bI_n) $$ 
The MLE of $\bPhi$ from $k$-th model is $\hat\bPhi_{(k)}= (\bZ'\bZ)^{-1}\bZ'\bY_k$. 
Thus, MLE of $\bPhi$ under $H_0$ is 
$$\hat\bPhi = \frac{1}{K}\sum_{k=1}^K (\bZ'\bZ)^{-1}\bZ'\bY_k$$
 The MLE of $\sig^2$ under $H_0$ is given by 
$$\hat\sig_0^2 = \frac{1}{nK}\sum_{k=1}^K (\bY_k - \bZ\hat\bPhi)' (\bY_k - \bZ\hat\bPhi)$$
The log-likelihood for the $k$-th genetic effect from the $k$-th marginal model is given by
$$l(\beta_k) \propto -\frac{1}{2\sig^2} (\bY_k - \beta_k \bX - \bZ \bPhi )'(\bY_k - \beta_k \bX - \bZ \bPhi )  $$
Marginal score for parameter $\beta_k$ under $H_{0k}$:
$$U_k = \dot{l}(\beta_k)\bigg|_{H_0}= \frac{1}{\sig^2} (\bY_k - \beta_k \bX - \bZ \bPhi )' \bX\bigg|_{H_0} = \frac{1}{\hat\sig_0^2} (\bY_k - \bZ\hat\bPhi)'\bX$$
Under the null, the variances and covariances of the marginal scores are:
$$ \Var(U_k) =  \frac{1}{\sig^4} \bX' \Var(\bY_k) \bX\bigg|_{H_0} = \frac{1}{\sig^2}\bX'\bX\bigg|_{H_0}$$
$$\Cov(U_k, U_j) = \frac{1}{\sig^4} \E(\bY_k'\bX \times \bY_j'\bX)\bigg|_{H_0} =
\frac{1}{\sig^4} \bX' \E(\bY_k \bY_j') \bX\bigg|_{H_0} = \frac{\rho}{\sig^2}\bX'\bX\bigg|_{H_0}    \: \forall \: j \neq k$$

\no Thus, under $H_0$, the score vector from the marginal normal model for $\bY$ is
$$\bU_M = \lt( \bY - \bs{1}' \otimes \bZ\hat\bPhi \rt)'\bX / \hat\sig_0^2$$
with covariance
$$\Cov(\bU_M) = \frac{1}{\sig^4} (\bX'\bX) \bSig \bigg|_{H_0} = \frac{1}{\hat\sig_0^4} (\bX'\bX) \hat\bSig_0
= \frac{1}{\hat\sig_0^4} (\bX'\bX) \frac{\lt( \bY - \bs{1}' \otimes \bZ\hat\bPhi \rt)'\lt( \bY - \bs{1}' \otimes \bZ\hat\bPhi \rt)}{n}$$

\no The SSU test based on the marginal normal score vector $\bU_M$ is 
$$T_{S} = \bU_M'\bU_M \overset{approx}\sim  a\chi^2_d + b$$
 where parameters $a$, $b$, $d$ are estimated as
$$a=\frac{\sum \del_i^3}{\sum \del_i^2}, \: \:
b=\sum \del_i  - \frac{(\sum \del_i^2)^2}{\sum \del_i^3}, \: \: 
d = \frac{(\sum \del_i^2)^3}{(\sum \del_i^3)^2}, \:\:
\{\del_i\}_{i=1}^K \text{ are the ordered eigenvalues of } \Cov(\bU_M)$$ \\

\newpage
\pagebreak
%%%%%%%%%  Web Table 1
\subsection{}	\label{app8}

\begin{center}
%\begin{longtable}{|c|c|c|c|c|c|c|c|}
{ \begin{footnotesize}
\begin{longtable}{|p{0.5cm} p{1.5cm} p{1.5cm} p{2cm} p{2cm} p{2cm} p{2cm} p{2cm} |}
\caption{List of all SNPs that exceed the significance of $0.05$ after Bonferroni correction for $2.5\times10^6$ SNPs (i.e., p-value threshold $2\times10^{-8}$) for the multivariate methods USAT and MANOVA. It is to be noted that most of these SNPs are in high linkage disequilibrium (LD). $p$ values for the univariate analysis of the individual traits are also provided. SNPs in bold are the ones detected solely by MANOVA but not by USAT. The abbreviations used are FG (Fasting Glucose), 2-hr GL (2-hour glucose  from an oral glucose tolerance test), FI (Fasting Insulin)} \label{t:sigall} \\
%\begin{tabular}{| c | c | c | c | c | c | c | c | }
\hline
  &  &  & MANOVA & USAT &  \multicolumn{3}{c}{Univariate Analysis $p$ }\\ \cline{6-8}
  chr & SNP & position & $p$ & $p$ & FG & 2-hr GL & FI \\ \hline
\endfirsthead
  {\ldots continued}\\ \hline
  &  &  & MANOVA & USAT &  \multicolumn{3}{c}{Univariate Analysis $p$ }\\ \cline{6-8}
  chr & SNP & position & $p$ & $p$ & FG & 2-hr GL & FI \\ \hline
\endhead
$2$ & $rs1260326$ & $27584444$ & $3.77\times10^{-15}$ & $4.44\times10^{-15}$ & $1.24\times10^{-4}$ & $6.26\times10^{-6}$ & $1.24\times10^{-5}$ \\
$2$ & $rs780094$ & $27594741$ & $9.99\times10^{-16}$ & $1.67\times10^{-15}$ & $7.34\times10^{-5}$ & $7.10\times10^{-6}$ & $4.65\times10^{-6}$ \\
$2$ & $rs780093$ & $27596107$ & $9.99\times10^{-16}$ & $1.67\times10^{-15}$ & $7.34\times10^{-5}$ & $7.10\times10^{-6}$ & $4.65\times10^{-6}$ \\
$2$ & $rs1260333$ & $27602128$ & $4.72\times10^{-11}$ & $8.14\times10^{-11}$ & $4.99\times10^{-4}$ & $3.84\times10^{-4}$ & $7.59\times10^{-5}$ \\
$2$ & $rs2911711$ & $27604050$ & $4.72\times10^{-11}$ & $8.14\times10^{-11}$ & $4.99\times10^{-4}$ & $3.84\times10^{-4}$ & $7.59\times10^{-5}$ \\
$2$ & $rs4665987$ & $27609329$ & $9.67\times10^{-10}$ & $2.31\times10^{-9}$ & $1.56\times10^{-2}$ & $4.49\times10^{-6}$ & $1.46\times10^{-2}$ \\
$2$ & $rs4665991$ & $27619788$ & $1.23\times10^{-9}$ & $2.57\times10^{-9}$ & $1.75\times10^{-2}$ & $4.83\times10^{-6}$ & $1.44\times10^{-2}$ \\
$2$ & $rs4665382$ & $27637305$ & $1.20\times10^{-9}$ & $2.54\times10^{-9}$ & $1.52\times10^{-2}$ & $5.13\times10^{-6}$ & $1.60\times10^{-2}$ \\
$2$ & $rs10208529$ & $27639692$ & $1.20\times10^{-9}$ & $2.54\times10^{-9}$ & $1.52\times10^{-2}$ & $5.13\times10^{-6}$ & $1.60\times10^{-2}$ \\
$2$ & $rs4665383$ & $27645059$ & $1.20\times10^{-9}$ & $2.54\times10^{-9}$ & $1.52\times10^{-2}$ & $5.13\times10^{-6}$ & $1.60\times10^{-2}$ \\
$2$ & $rs1919127$ & $27654997$ & $1.20\times10^{-9}$ & $2.54\times10^{-9}$ & $1.52\times10^{-2}$ & $5.13\times10^{-6}$ & $1.60\times10^{-2}$ \\
$2$ & $rs1919128$ & $27655263$ & $1.20\times10^{-9}$ & $2.54\times10^{-9}$ & $1.52\times10^{-2}$ & $5.13\times10^{-6}$ & $1.60\times10^{-2}$ \\
$2$ & $rs12478841$ & $27665226$ & $1.06\times10^{-9}$ & $2.40\times10^{-9}$ & $1.63\times10^{-2}$ & $4.96\times10^{-6}$ & $1.31\times10^{-2}$ \\
$2$ & $rs6760250$ & $27665756$ & $9.94\times10^{-10}$ & $2.34\times10^{-9}$ & $1.49\times10^{-2}$ & $6.01\times10^{-6}$ & $1.07\times10^{-2}$ \\
$2$ & $rs13022873$ & $27669014$ & $9.94\times10^{-10}$ & $2.34\times10^{-9}$ & $1.49\times10^{-2}$ & $6.01\times10^{-6}$ & $1.07\times10^{-2}$ \\
$2$ & $rs12467476$ & $27679219$ & $9.86\times10^{-10}$ & $2.33\times10^{-9}$ & $1.53\times10^{-2}$ & $6.00\times10^{-6}$ & $1.02\times10^{-2}$ \\
$2$ & $rs2384656$ & $27685559$ & $9.86\times10^{-10}$ & $2.33\times10^{-9}$ & $1.53\times10^{-2}$ & $6.00\times10^{-6}$ & $1.02\times10^{-2}$ \\
$2$ & $rs4666002$ & $27694144$ & $8.64\times10^{-10}$ & $1.43\times10^{-9}$ & $1.61\times10^{-2}$ & $5.56\times10^{-6}$ & $9.02\times10^{-3}$ \\
$2$ & $rs3749147$ & $27705422$ & $6.52\times10^{-9}$ & $1.31\times10^{-8}$ & $3.65\times10^{-2}$ & $6.56\times10^{-6}$ & $1.86\times10^{-2}$ \\
$2$ & $rs13002853$ & $27706749$ & $6.52\times10^{-9}$ & $1.31\times10^{-8}$ & $3.65\times10^{-2}$ & $6.56\times10^{-6}$ & $1.86\times10^{-2}$ \\
$2$ & $rs13431652$ & $169461661$ & $1.85\times10^{-13}$ & $5.48\times10^{-13}$ & $2.24\times10^{-12}$ & $9.57\times10^{-1}$ & $2.85\times10^{-1}$ \\
$2$ & $rs1402837$ & $169465600$ & $4.91\times10^{-9}$ & $1.15\times10^{-8}$ & $2.78\times10^{-10}$ & $5.18\times10^{-2}$ & $8.07\times10^{-1}$ \\
$2$ & $rs573225$ & $169465787$ & $4.55\times10^{-14}$ & $5.81\times10^{-14}$ & $9.75\times10^{-13}$ & $9.83\times10^{-1}$ & $2.33\times10^{-1}$ \\
$2$ & $rs560887$ & $169471394$ & $5.55\times10^{-16}$ & $1.33\times10^{-15}$ & $1.24\times10^{-14}$ & $8.87\times10^{-1}$ & $2.93\times10^{-1}$ \\
$2$ & $rs563694$ & $169482317$ & $1.54\times10^{-14}$ & $2.91\times10^{-14}$ & $4.12\times10^{-14}$ & $3.50\times10^{-1}$ & $3.54\times10^{-1}$ \\
$2$ & $rs537183$ & $169482892$ & $1.54\times10^{-14}$ & $2.91\times10^{-14}$ & $4.12\times10^{-14}$ & $3.50\times10^{-1}$ & $3.54\times10^{-1}$ \\
$2$ & $rs502570$ & $169483205$ & $1.54\times10^{-14}$ & $2.91\times10^{-14}$ & $4.12\times10^{-14}$ & $3.50\times10^{-1}$ & $3.54\times10^{-1}$ \\
$2$ & $rs475612$ & $169484992$ & $3.54\times10^{-13}$ & $7.12\times10^{-13}$ & $3.74\times10^{-13}$ & $3.63\times10^{-1}$ & $5.12\times10^{-1}$ \\
$2$ & $rs557462$ & $169485841$ & $1.54\times10^{-14}$ & $2.91\times10^{-14}$ & $4.12\times10^{-14}$ & $3.50\times10^{-1}$ & $3.54\times10^{-1}$ \\
$2$ & $rs478333$ & $169487402$ & $8.61\times10^{-10}$ & $1.42\times10^{-9}$ & $3.33\times10^{-10}$ & $4.16\times10^{-1}$ & $6.41\times10^{-1}$ \\
$2$ & $rs496550$ & $169487958$ & $8.61\times10^{-10}$ & $1.42\times10^{-9}$ & $3.33\times10^{-10}$ & $4.16\times10^{-1}$ & $6.41\times10^{-1}$ \\
$2$ & $rs473351$ & $169488142$ & $2.58\times10^{-11}$ & $6.05\times10^{-11}$ & $1.38\times10^{-11}$ & $2.75\times10^{-1}$ & $5.29\times10^{-1}$ \\
$2$ & $rs575671$ & $169489064$ & $2.58\times10^{-11}$ & $6.05\times10^{-11}$ & $1.38\times10^{-11}$ & $2.75\times10^{-1}$ & $5.29\times10^{-1}$ \\
$2$ & $rs519887$ & $169489131$ & $8.50\times10^{-10}$ & $1.41\times10^{-9}$ & $3.45\times10^{-10}$ & $4.40\times10^{-1}$ & $6.41\times10^{-1}$ \\
$2$ & $rs486981$ & $169490395$ & $2.72\times10^{-14}$ & $4.05\times10^{-14}$ & $1.15\times10^{-13}$ & $6.17\times10^{-1}$ & $3.89\times10^{-1}$ \\
$2$ & $rs484066$ & $169490727$ & $2.01\times10^{-12}$ & $2.32\times10^{-12}$ & $6.10\times10^{-12}$ & $8.54\times10^{-1}$ & $4.87\times10^{-1}$ \\
$2$ & $rs569805$ & $169491126$ & $2.72\times10^{-14}$ & $4.05\times10^{-14}$ & $1.15\times10^{-13}$ & $6.17\times10^{-1}$ & $3.89\times10^{-1}$ \\
$2$ & $rs579060$ & $169491285$ & $2.45\times10^{-14}$ & $3.77\times10^{-14}$ & $9.95\times10^{-14}$ & $6.04\times10^{-1}$ & $3.93\times10^{-1}$ \\
$2$ & $rs17540154$ & $169492739$ & $3.46\times10^{-9}$ & $6.48\times10^{-9}$ & $3.41\times10^{-10}$ & $2.33\times10^{-1}$ & $9.19\times10^{-1}$ \\
$2$ & $rs508506$ & $169493201$ & $3.55\times10^{-14}$ & $4.86\times10^{-14}$ & $8.64\times10^{-14}$ & $5.74\times10^{-1}$ & $4.97\times10^{-1}$ \\
$2$ & $rs503931$ & $169493695$ & $8.50\times10^{-10}$ & $1.41\times10^{-9}$ & $3.45\times10^{-10}$ & $4.40\times10^{-1}$ & $6.41\times10^{-1}$ \\
$2$ & $rs551754$ & $169495932$ & $8.50\times10^{-10}$ & $1.41\times10^{-9}$ & $3.45\times10^{-10}$ & $4.40\times10^{-1}$ & $6.41\times10^{-1}$ \\
$2$ & $rs497692$ & $169497262$ & $8.41\times10^{-10}$ & $1.41\times10^{-9}$ & $3.27\times10^{-10}$ & $4.24\times10^{-1}$ & $6.46\times10^{-1}$ \\
$2$ & $rs494874$ & $169497552$ & $1.42\times10^{-13}$ & $5.06\times10^{-13}$ & $1.70\times10^{-13}$ & $5.38\times10^{-1}$ & $6.48\times10^{-1}$ \\
$2$ & $rs552976$ & $169499684$ & $1.55\times10^{-13}$ & $5.19\times10^{-13}$ & $1.87\times10^{-13}$ & $5.31\times10^{-1}$ & $6.37\times10^{-1}$ \\
$2$ & $\bs{rs472614}$ & $169500667$ & $1.26\times10^{-8}$ & $2.65\times10^{-8}$ & $3.55\times10^{-9}$ & $5.72\times10^{-1}$ & $8.17\times10^{-1}$ \\
$2$ & $rs565412$ & $169502529$ & $9.14\times10^{-9}$ & $1.57\times10^{-8}$ & $4.16\times10^{-9}$ & $7.18\times10^{-1}$ & $7.34\times10^{-1}$ \\
$2$ & $rs567074$ & $169502677$ & $3.45\times10^{-10}$ & $5.76\times10^{-10}$ & $1.51\times10^{-10}$ & $6.13\times10^{-1}$ & $8.04\times10^{-1}$ \\
$2$ & $rs479682$ & $169502933$ & $7.13\times10^{-9}$ & $1.37\times10^{-8}$ & $3.33\times10^{-9}$ & $6.89\times10^{-1}$ & $7.06\times10^{-1}$ \\
$2$ & $rs480562$ & $169503017$ & $7.46\times10^{-9}$ & $1.40\times10^{-8}$ & $3.43\times10^{-9}$ & $6.84\times10^{-1}$ & $7.07\times10^{-1}$ \\
$2$ & $rs2685803$ & $169504531$ & $7.46\times10^{-9}$ & $1.40\times10^{-8}$ & $3.43\times10^{-9}$ & $6.84\times10^{-1}$ & $7.07\times10^{-1}$ \\
$2$ & $rs2544367$ & $169504534$ & $4.54\times10^{-9}$ & $7.54\times10^{-9}$ & $2.46\times10^{-9}$ & $7.19\times10^{-1}$ & $6.85\times10^{-1}$ \\
$2$ & $rs2685805$ & $169505306$ & $4.54\times10^{-9}$ & $7.54\times10^{-9}$ & $2.46\times10^{-9}$ & $7.19\times10^{-1}$ & $6.85\times10^{-1}$ \\
$2$ & $rs1581397$ & $169505898$ & $4.05\times10^{-9}$ & $7.05\times10^{-9}$ & $2.40\times10^{-9}$ & $7.37\times10^{-1}$ & $6.66\times10^{-1}$ \\
$2$ & $rs2685814$ & $169506865$ & $3.62\times10^{-9}$ & $6.63\times10^{-9}$ & $2.19\times10^{-9}$ & $7.48\times10^{-1}$ & $6.71\times10^{-1}$ \\
$2$ & $rs853789$ & $169509734$ & $2.00\times10^{-15}$ & $2.66\times10^{-15}$ & $8.50\times10^{-15}$ & $6.36\times10^{-1}$ & $4.84\times10^{-1}$ \\
$2$ & $rs860510$ & $169509874$ & $3.62\times10^{-9}$ & $6.63\times10^{-9}$ & $2.19\times10^{-9}$ & $7.48\times10^{-1}$ & $6.71\times10^{-1}$ \\
$2$ & $rs853788$ & $169510151$ & $3.62\times10^{-9}$ & $6.63\times10^{-9}$ & $2.19\times10^{-9}$ & $7.48\times10^{-1}$ & $6.71\times10^{-1}$ \\
$2$ & $rs853787$ & $169510498$ & $2.00\times10^{-15}$ & $2.66\times10^{-15}$ & $8.50\times10^{-15}$ & $6.36\times10^{-1}$ & $4.84\times10^{-1}$ \\
$2$ & $rs853786$ & $169510556$ & $3.62\times10^{-9}$ & $6.63\times10^{-9}$ & $2.19\times10^{-9}$ & $7.48\times10^{-1}$ & $6.71\times10^{-1}$ \\
$2$ & $rs862662$ & $169510575$ & $1.52\times10^{-10}$ & $2.42\times10^{-10}$ & $1.01\times10^{-10}$ & $6.80\times10^{-1}$ & $7.17\times10^{-1}$ \\
$2$ & $rs853785$ & $169510840$ & $3.62\times10^{-9}$ & $6.63\times10^{-9}$ & $2.19\times10^{-9}$ & $7.48\times10^{-1}$ & $6.71\times10^{-1}$ \\
$2$ & $rs853784$ & $169511920$ & $5.48\times10^{-9}$ & $1.21\times10^{-8}$ & $3.37\times10^{-9}$ & $8.18\times10^{-1}$ & $7.11\times10^{-1}$ \\
$2$ & $rs853783$ & $169513757$ & $5.48\times10^{-9}$ & $1.21\times10^{-8}$ & $3.37\times10^{-9}$ & $8.18\times10^{-1}$ & $7.11\times10^{-1}$ \\
$2$ & $rs853781$ & $169514567$ & $3.43\times10^{-10}$ & $5.74\times10^{-10}$ & $2.19\times10^{-10}$ & $7.37\times10^{-1}$ & $7.53\times10^{-1}$ \\
$2$ & $rs853780$ & $169515728$ & $7.62\times10^{-9}$ & $1.42\times10^{-8}$ & $4.68\times10^{-9}$ & $7.98\times10^{-1}$ & $6.76\times10^{-1}$ \\
$2$ & $rs1101533$ & $169516768$ & $7.62\times10^{-9}$ & $1.42\times10^{-8}$ & $4.68\times10^{-9}$ & $7.98\times10^{-1}$ & $6.76\times10^{-1}$ \\
$2$ & $rs853779$ & $169517918$ & $4.39\times10^{-9}$ & $7.39\times10^{-9}$ & $2.88\times10^{-9}$ & $7.91\times10^{-1}$ & $6.67\times10^{-1}$ \\
$2$ & $rs853778$ & $169519470$ & $1.28\times10^{-9}$ & $2.62\times10^{-9}$ & $1.52\times10^{-9}$ & $9.11\times10^{-1}$ & $5.79\times10^{-1}$ \\
$2$ & $rs853773$ & $169522593$ & $1.63\times10^{-9}$ & $2.96\times10^{-9}$ & $2.44\times10^{-9}$ & $8.12\times10^{-1}$ & $7.64\times10^{-1}$ \\
$2$ & $rs16844037$ & $210561151$ & $1.92\times10^{-9}$ & $1.04\times10^{-9}$ & $2.16\times10^{-1}$ & $2.56\times10^{-2}$ & $1.14\times10^{-10}$ \\
$6$ & $rs12154183$ & $295724$ & $2.81\times10^{-9}$ & $1.02\times10^{-9}$ & $4.01\times10^{-1}$ & $9.55\times10^{-1}$ & $1.21\times10^{-9}$ \\
$15$ & $\bs{rs7170293}$ & $60023665$ & $1.98\times10^{-8}$ & $6.28\times10^{-8}$ & $5.77\times10^{-3}$ & $3.20\times10^{-5}$ & $2.68\times10^{-1}$ \\
$15$ & $\bs{rs7166891}$ & $60026596$ & $1.98\times10^{-8}$ & $6.28\times10^{-8}$ & $5.77\times10^{-3}$ & $3.20\times10^{-5}$ & $2.68\times10^{-1}$ \\
$15$ & $\bs{rs7172145}$ & $60026989$ & $1.98\times10^{-8}$ & $6.28\times10^{-8}$ & $5.77\times10^{-3}$ & $3.20\times10^{-5}$ & $2.68\times10^{-1}$ \\
$15$ & $\bs{rs4587915}$ & $60029254$ & $1.44\times10^{-8}$ & $2.82\times10^{-8}$ & $9.60\times10^{-3}$ & $1.12\times10^{-5}$ & $3.28\times10^{-1}$ \\
$15$ & $\bs{rs8034335}$ & $60074748$ & $1.19\times10^{-8}$ & $2.59\times10^{-8}$ & $1.43\times10^{-2}$ & $5.58\times10^{-6}$ & $3.35\times10^{-1}$ \\
$15$ & $\bs{rs8034216}$ & $60074820$ & $1.19\times10^{-8}$ & $2.59\times10^{-8}$ & $1.43\times10^{-2}$ & $5.58\times10^{-6}$ & $3.35\times10^{-1}$ \\
$15$ & $rs17271305$ & $60120272$ & $6.87\times10^{-9}$ & $1.35\times10^{-8}$ & $6.29\times10^{-3}$ & $1.17\times10^{-5}$ & $2.86\times10^{-1}$ \\
$15$ & $rs17271340$ & $60135177$ & $8.99\times10^{-9}$ & $1.55\times10^{-8}$ & $1.68\times10^{-2}$ & $3.71\times10^{-6}$ & $3.08\times10^{-1}$ \\
$15$ & $rs8039105$ & $60146377$ & $8.71\times10^{-9}$ & $1.53\times10^{-8}$ & $1.65\times10^{-2}$ & $3.75\times10^{-6}$ & $3.05\times10^{-1}$ \\
$15$ & $\bs{rs7163757}$ & $60178900$ & $1.68\times10^{-8}$ & $5.98\times10^{-8}$ & $7.98\times10^{-4}$ & $3.52\times10^{-4}$ & $4.88\times10^{-2}$ \\
$15$ & $rs8037894$ & $60181556$ & $8.20\times10^{-9}$ & $1.48\times10^{-8}$ & $4.09\times10^{-4}$ & $4.89\times10^{-4}$ & $3.12\times10^{-2}$ \\
$15$ & $\bs{rs6494307}$ & $60181982$ & $1.68\times10^{-8}$ & $5.98\times10^{-8}$ & $7.98\times10^{-4}$ & $3.52\times10^{-4}$ & $4.88\times10^{-2}$ \\
$15$ & $\bs{rs7167878}$ & $60183481$ & $1.68\times10^{-8}$ & $5.98\times10^{-8}$ & $7.98\times10^{-4}$ & $3.52\times10^{-4}$ & $4.88\times10^{-2}$ \\
$15$ & $\bs{rs7172432}$ & $60183681$ & $1.68\times10^{-8}$ & $5.98\times10^{-8}$ & $7.98\times10^{-4}$ & $3.52\times10^{-4}$ & $4.88\times10^{-2}$ \\ \hline
%\end{tabular}
\end{longtable}
\end{footnotesize} }
\end{center}

\newpage
\pagebreak
%%%%%%%%%  Web Figure 1
\subsection{} 	\label{app9}

\begin{figure}[b!]
	\begin{center}
	\caption[Simulation 1 power curves: 2 traits, opposite direction effects]{Empirical power curves of the different existing association tests for $K=2$ traits and different within trait correlation values $\rho=-0.8,-0.6,-0.4,-0.2,0.2,...,0.8$ based on $N=500$ datasets with $n=4,000$ unrelated subjects. Opposite direction but same size genetic effect used when both traits are associated (i.e., datasets are generated from an alternative model $H_{a2,2}:\beta_1=-\beta_2>0$). Effect size of $0.25$ is used for the associated traits. The power is plotted along y-axis while the fraction of traits associated with the genetic variant is plotted along x-axis.}
	\includegraphics[height=6.5in]{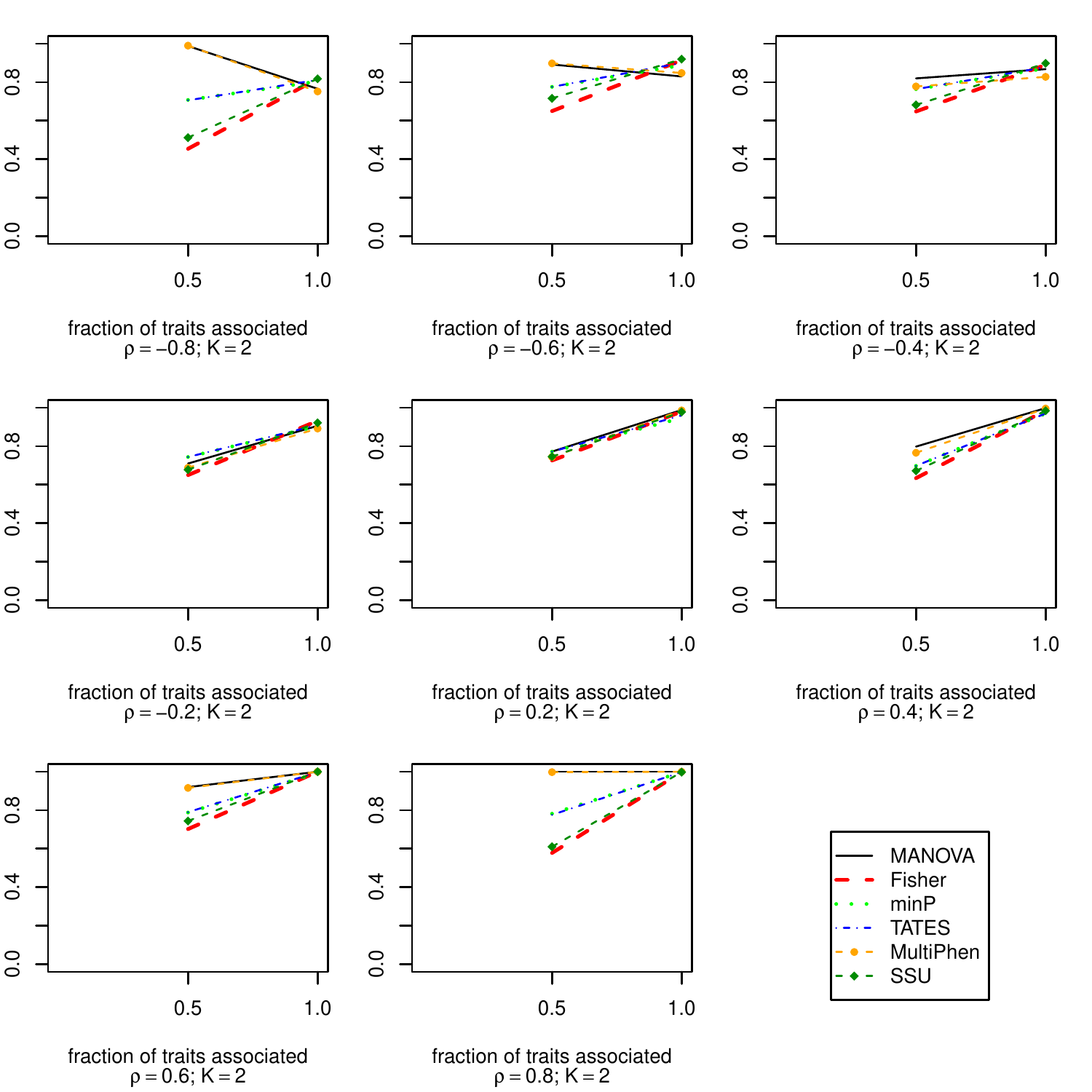}
	\label{fig2}
	\end{center}
\end{figure}

\newpage
\pagebreak
%%%%%%%%%  Web Figure
\subsection{}	\label{app10}

\begin{figure}[b!]
	\begin{center}
	\caption[Comparison of USAT with SSU and MANOVA: $K=5,10,20$ traits, SNP at m.a.f. $0.05$]{Empirical power curves of the SSU and MANOVA tests along with our novel approach USAT. $K=5,10,20$ traits have been simulated at different within trait correlation values $\rho=0.2,0.4,0.6$. For each value of $K$ and $\rho$, there were $N=500$ datasets of $n=400$ unrelated individuals. A single SNP with minor allele frequency (m.a.f.) $0.05$ was simulated. Same effect size of $0.725$ was used for the traits that are associated. The power is plotted along y-axis while the fraction of traits associated is plotted along x-axis. 
This figure shows that the  relative behavior of MANOVA and the SSU test does not vary much with change in m.a.f. Since our proposed test USAT is derived from an optimal weighted combination of MANOVA and the SSU test, the performance of USAT compared to MANOVA or SSU also does not vary much with change in m.a.f.}
	\includegraphics[height=6.8in]{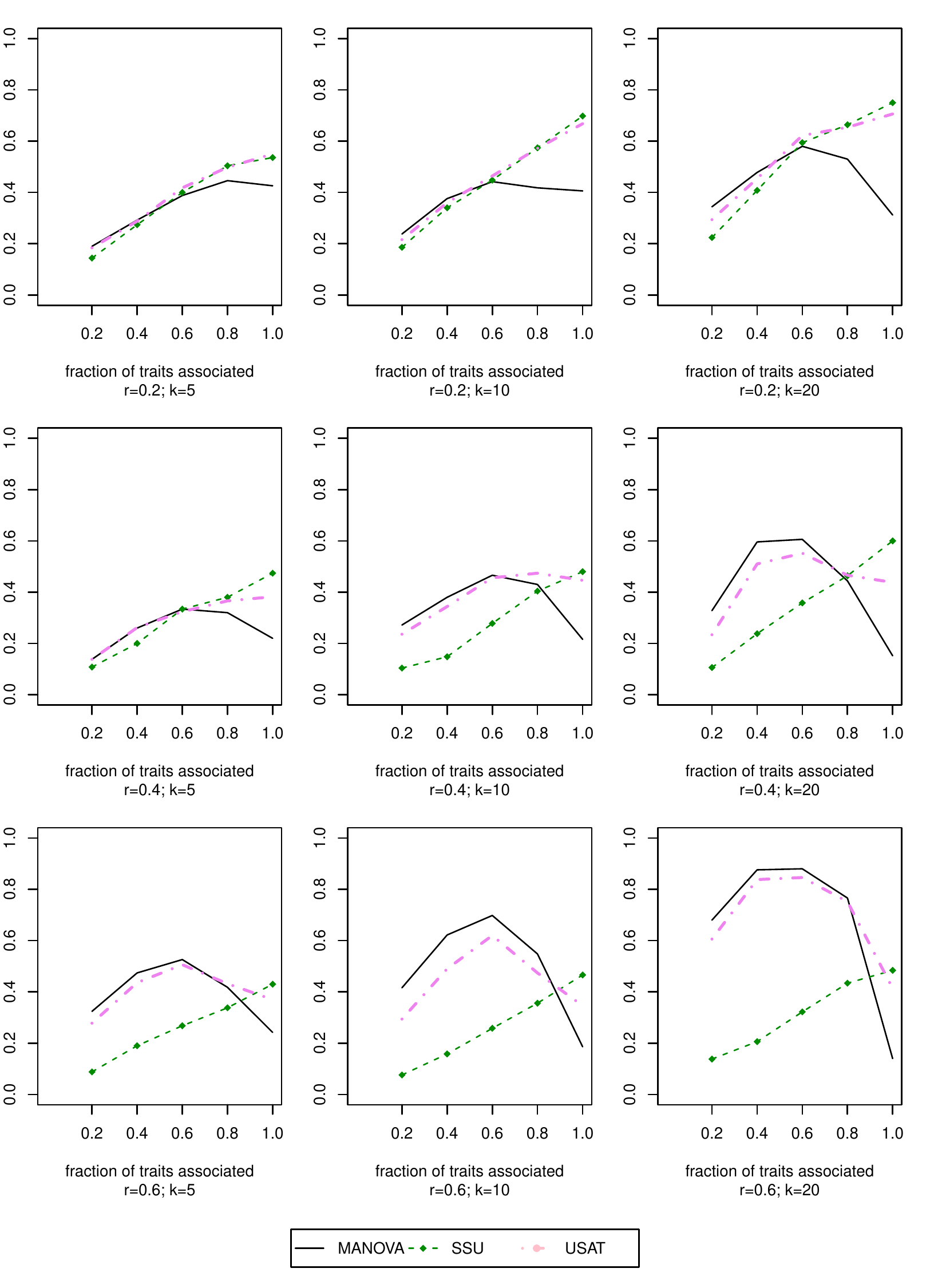}
	\label{fig4-maf5}
	\end{center}
\end{figure}